\documentclass[preprint, 11pt]{aastex}
\usepackage{amsmath}	
\usepackage{natbib}
\usepackage[margin=20mm]{geometry}
\usepackage{graphicx}
\usepackage{grffile}

\newcommand{\Pvec}{\boldsymbol P}
\newcommand{\Qvec}{\boldsymbol Q}
\newcommand{\fvec}{\boldsymbol f}

 \newcommand{\rpc}{r_{\rm pc}}
 \newcommand{\rlc}{r_{\rm lc}}
 
 \newcommand{\Rvec}{\boldsymbol R}
\newcommand{\Svec}{\boldsymbol S}
\newcommand{\Uvec}{\boldsymbol U}
\newcommand{\Ivec}{\boldsymbol I}

\newcommand{\Evec}{\boldsymbol E}
\newcommand{\Bvec}{\boldsymbol B}
\newcommand{\Omegavec}{\boldsymbol \Omega}
\newcommand{\Jvec}{\boldsymbol J}
\newcommand{\rhoGJ}{\rho_{\rm GJ}}
\newcommand{\NF}{f_n}
\newcommand{\Pf}{P_{\rm f}}
\newcommand{\Porb}{P_{\rm orb}}
\newcommand{\Msun}{M_{\odot}}

\newcommand{\be}{\begin{eqnarray}}
\newcommand{\ee}{\end{eqnarray}}

\newcommand{\Tiave}{T_i}
\newcommand{\Toneave}{T_1}
\newcommand{\Ttwoave}{T_2}
\newcommand{\Tthreeave}{T_3}

\newcommand{\Tbar}{\overline{T}}
\newcommand{\snr}{\rm S/N}

\begin{document}
\singlespace

\title{Pulsar State Switching from Markov Transitions and Stochastic Resonance}
\author{J. M. Cordes }
\affil{Astronomy Department, Cornell University, Ithaca, NY 14853} 
\email{cordes@astro.cornell.edu}
\author{\today}

\begin{abstract}
Markov processes are shown to be consistent with metastable states
seen in pulsar phenomena, including intensity nulling, pulse-shape mode changes, subpulse drift rates, spindown rates, and X-ray emission,
based on the typically broad and monotonic distributions of state lifetimes. Markovianity implies a nonlinear magnetospheric system in which  state changes occur stochastically, corresponding to transitions between local minima in an effective potential. 
State durations (though not transition times) are thus largely decoupled from the characteristic time scales of various magnetospheric processes. 
Dyadic states are common but some objects show at least four states with some transitions forbidden.      Another case is
the long-term intermittent pulsar B1931+24 that has binary
radio-emission and torque states with wide, but
{\em non}-monotonic duration distributions. It also shows a quasi-period  of $38\pm5$ days  in a 13-yr time sequence,  suggesting stochastic resonance in a Markov system with a forcing function that  could be strictly periodic or quasi-periodic.  Nonlinear phenomena are associated with time-dependent activity in the acceleration region near each magnetic polar cap.  
The polar-cap diode is altered by feedback from the outer magnetosphere and by return currents from an equatorial disk that may also cause the neutron star to episodically charge and discharge.  Orbital perturbations in the disk provide a natural periodicity for the forcing function in the stochastic resonance interpretation of B1931+24.  Disk dynamics may introduce additional time scales in observed phenomena.   Future work can  test the Markov interpretation,  identify which pulsar types have a propensity for state changes, and clarify the role of selection effects.
\end{abstract}

\section{Introduction}
\label{sec:intro}

Ever since their discovery it has been well known that  radio pulses from pulsars
 vary dramatically on a wide range of time scales.  
By comparison, incoherent optical to gamma-ray emission shows little variability, apart from the basic pulsation.  Consequently,   radio variability was widely considered to  result from changes in radio coherence that did not couple to high-energy emission or to the overall loss of rotational energy.     That has now changed with the identification of large changes
in spindown rates on time scales $\sim 10^6$~s 
\citep[][]{2006Sci...312..549K,2012ApJ...746...63C,2012ApJ...758..141L} that correlate with distinct on-and-off
radio emission states.  Equally important is the detection of 
 X-ray emission state changes  
 on time scales
$\sim 10^3$-$10^4$~s   that are linked to specific radio modes
in the pulsar B0943+10 \citep{h+13}. 
These phenomena indicate that
 radio emission, though  energetically small, traces  the dominant energy channels of 
 the magnetosphere, including  acceleration
regions and the large scale structure of the magnetosphere itself.

The {\em ansatz} for this paper is consistent with similar ideas expressed
by other authors 
\citep[][]{
1975ApJ...196...51R,
1979SSRv...24..567C, 
1982MNRAS.200.1081J,
1983AIPC..101...98C,
1986MNRAS.222..577J,
2010MNRAS.408L..41T, 
2011MNRAS.414..759J, 
2012ApJ...746L..24L, 
2012ApJ...752..155V}, 
 that the observed state changes on both short and 
long time scales reflect changes in 
(1)  the global state and energetics of a pulsar's magnetosphere, as
implied by the large changes in spindown rate seen in intermittent pulsars; 
(2) the accelerating potential directly above the magnetic polar caps that generates
relativistic particles, as evidenced by mode changes; and
(3) the temperature of the hot polar cap due to particle bombardment as demonstrated
by X-ray low and high states that correlate with mode changes.    

This paper  addresses issues in understanding pulsar magnetospheres using the statistical framework of Markov chains: 
 (1)  What is an accurate yet parsimonious description of state changes in pulse sequences?
 (2)  Are some or even all observed state changes connected through some common
       paradigm? In particular, do intensity nulls occurring on time scales of hours or less
       have the same physical origin as the much longer intermittency on days to week-like
       time scales?
 (3) How directly related are  observed phenomena to  physical changes 
       in the pulsar magnetosphere?  Are there  underlying state changes that are not directly
       observable?
 (4) What determines the occurrence of a state change?   Is it triggered externally or is it
       a threshold effect in the internal dynamics of the magnetosphere?

In Section~\ref{sec:phenomena} we discuss observations of state changes
that are relevant to our modeling.    
In Sections~\ref{sec:markov}-\ref{sec:SR} we summarize general properties
of Markov processes and apply them to some of the key observations.   
In Sections~\ref{sec:timing} and \ref{sec:B0823+26} we discuss  spin variations expected from a two-state Markov process with reference to the  
medium-term intermittent pulsar B0823+26 and the long-term intermittent pulsar B1931+24.
In Section~\ref{sec:states}   we relate our findings to pulsar magnetospheres and their surroundings and in Section~\ref{sec:summary} we summarize our results and
our conclusions.

\section{State Changes in Pulsars}
\label{sec:phenomena}

Table~\ref{tab:phenomena} itemizes some of the phenomena that we discuss,
all of which  have been seen only in relatively
old, canonical pulsars but not in young pulsars,   in millisecond pulsars (MSPs),
or magnetars.    This suggests that
state changes occur in objects that are in a particular region of 
magnetic field - spin-period space, though it
is possible that selection effects, such as the inability to study single pulses from
most MSPs, prohibits recognition of state changes.   This situation will change as
more sensitive measurements are obtained.

\begin{figure}[h!]
\begin{center}
\includegraphics[scale=0.70, angle=0]{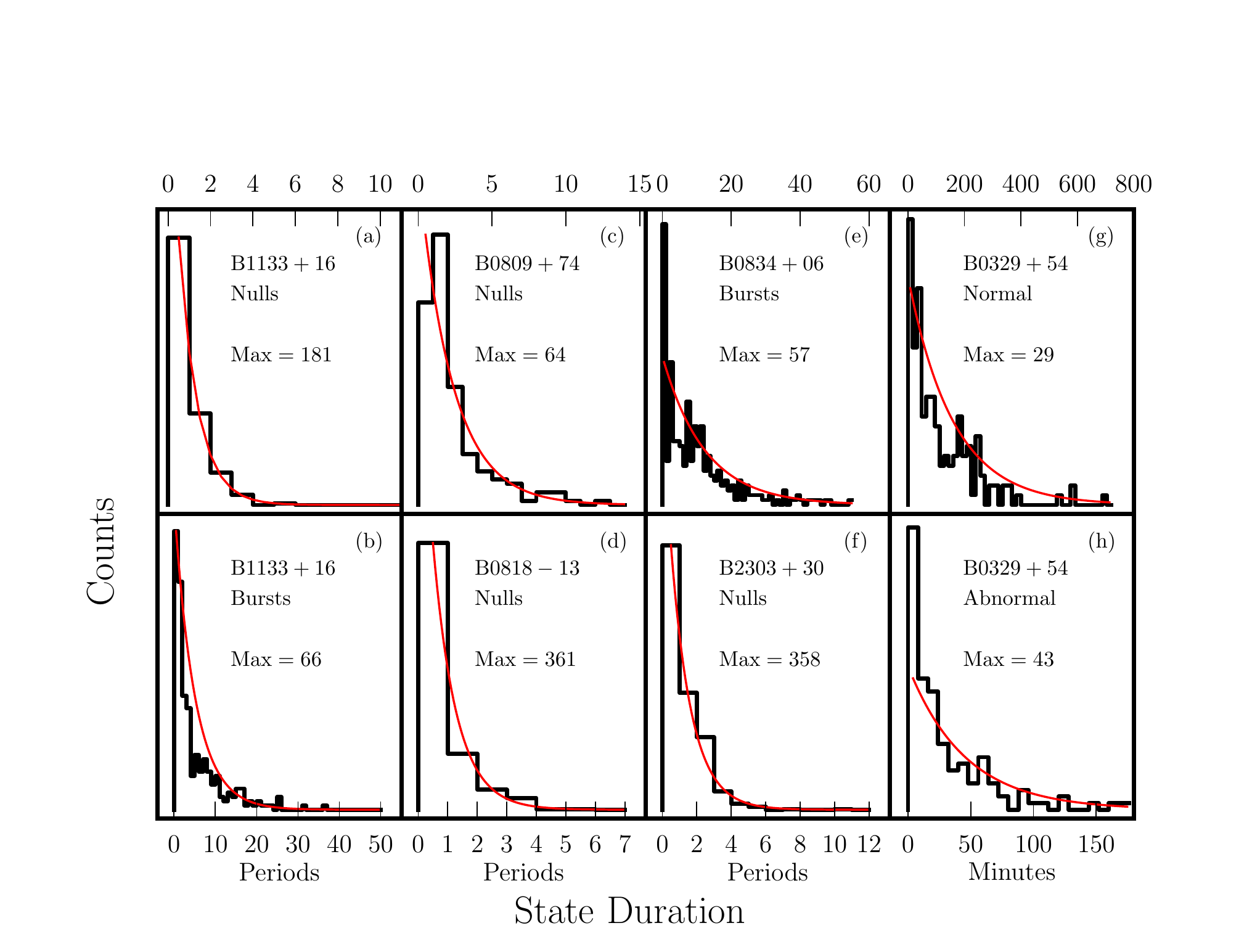}
\caption{Histograms of durations of nulls, bursts, and modes for six pulsars.
Frames (a) and (b) show null and burst histograms for B1133+16 at 325~MHz from
\citet[][]{2007A&A...462..257B};
(c) Nulls for B0809+74 at 350~MHz \citep[][]{2002A&A...387..169V};
(d) Nulls for B0818-13 \citep[][]{2004A&A...425..255J};
(e) Bursts for B0834+06 \citep[][]{2009MNRAS.395.1529R};
(f)  Nulls for B2303+30 \citep[][]{2005MNRAS.357..859R};
(g) and (h) Normal and abnormal mode durations  for B0329+54
\citep[][]{2011ApJ...741...48C}.
The smooth lines are  Equation~\ref{eq:pdf} evaluated using the mean state duration
and the number of counts in each case; they are not least-squares fits and are given solely
as fiducial examples.  The maximum number of counts is given in each frame.
\label{fig:fig1}
}
\end{center}
\end{figure}

\begin{figure}[h!]
\begin{center}
\includegraphics[scale=0.40, angle=0]{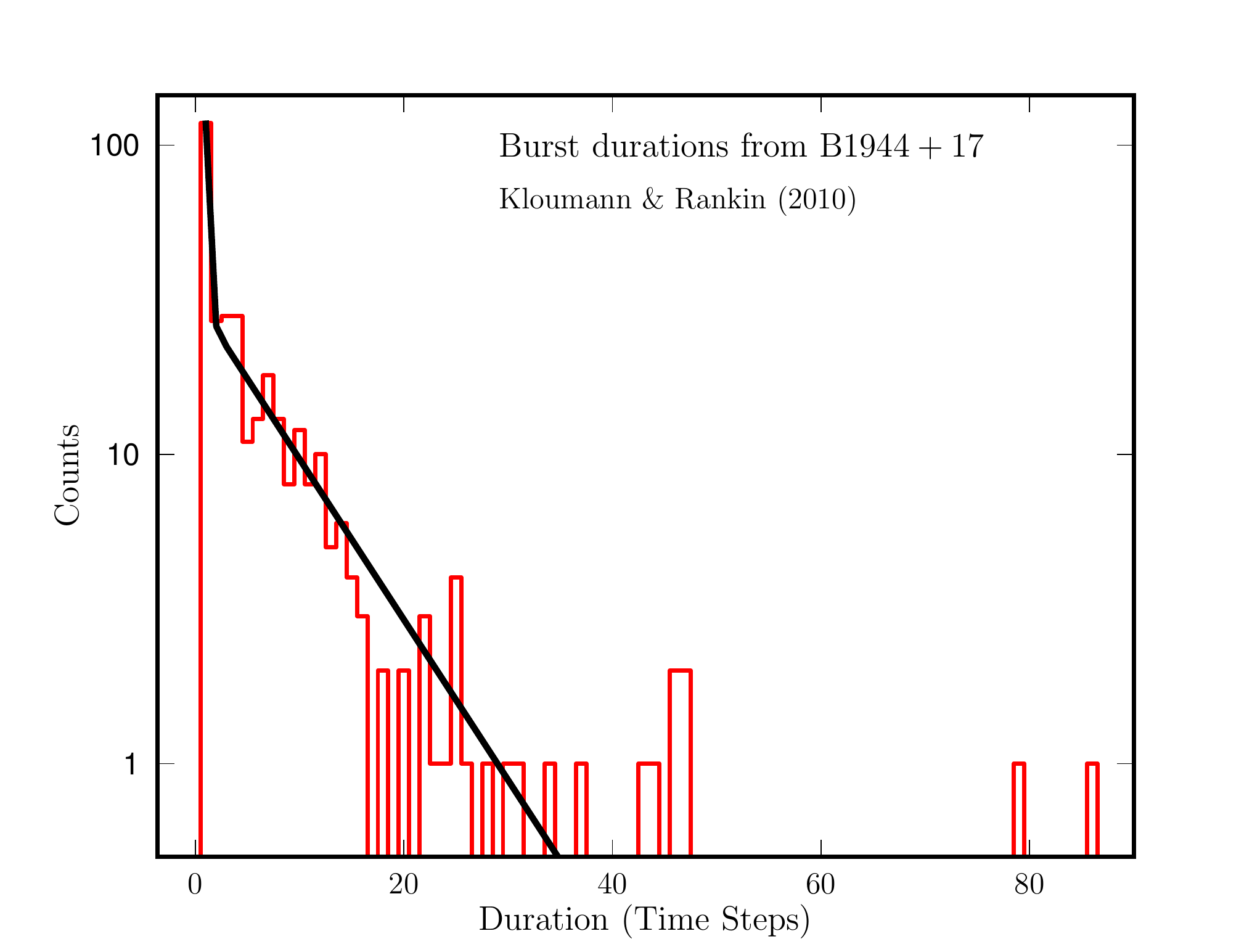}
\includegraphics[scale=0.40, angle=0]{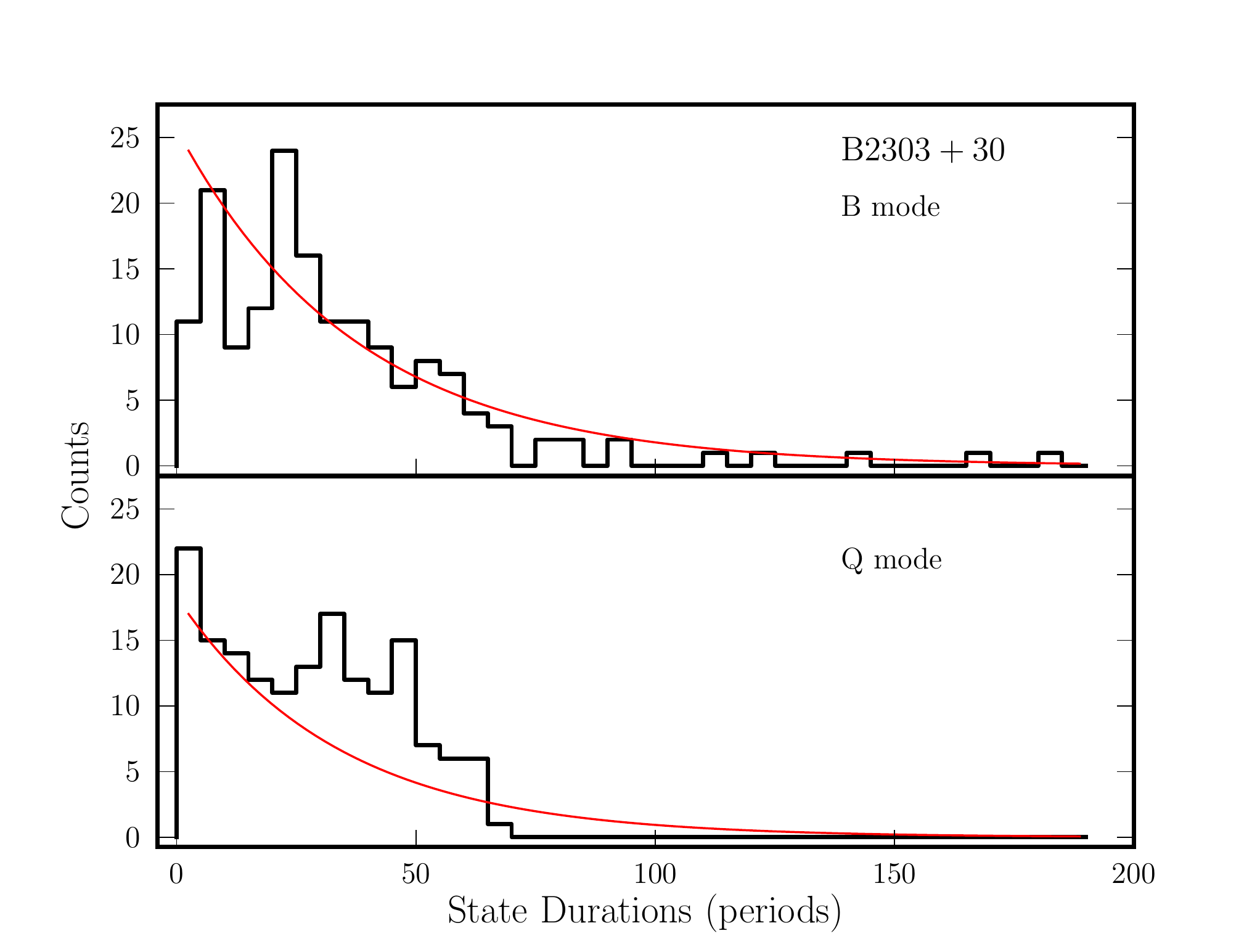}
\caption{
 Left: Histogram of burst durations for B1944+17 at 327 MHz 
(after Kloumann \& Rankin 2010).   The solid line shows the predicted distribution
for a four-state Markov model where two burst states are  counted together
as discussed in the text (Section~\ref{sec:duals}).
Right: Histograms of the durations of the burst (B) and quiet (Q) modes for
B2303+30 from \citet[][]{2005MNRAS.357..859R}.
The smooth lines are  Equation~\ref{eq:pdf} evaluated using the mean state duration
and the number of counts in each case. 
\label{fig:fig2}
}
\end{center}
\end{figure}

The distribution of  state durations is one of the primary tools used in this paper
for characterizing state changes. 
Figure~\ref{fig:fig1} presents examples  for six pulsars that show nulls and bursts
or pulse-shape mode changes.  Within counting errors, the histograms decrease
monotonically as expected for  a two-state Markov process (viz. Equation~\ref{eq:pdf} , as discussed further below). 
In several cases (B0834+06, the abnormal mode for B0329+54, 
and bursts for B1133+16)
there is an excess of short-duration states, as also pointed out  for other objects \citep[][]{2002A&A...387..169V,2010MNRAS.408...40K,2012MNRAS.424.1197G}.
The duration histogram for bursts is shown for one of those, B1944+17,
in Figure~\ref{fig:fig2} (left panel). 
Apparent counterexamples  to these trends are shown in the right-hand panel
of Figure~\ref{fig:fig2},
which gives duration histograms of bursts (B) and quiet (Q) modes from B2303+30.  
However, as pointed out
by \citet[][]{2005MNRAS.357..859R}, whose Figure~4 is the basis for
Figure~\ref{fig:fig2}, short-duration B and Q sequences 
are probably undercounted, causing some of the departure from an exponential-like
distribution. 

While an excess of short-duration states appears to be intrinsic to 
pulsar emission in some cases,  in others there is
inadequate signal-to-noise ratio (\snr) to avoid
false transitions between nulls and bursts caused by additive noise.  Since noise is independent
between pulse periods, the durations of such mis-identified states are typically  only one period long.   This is discussed further in Section~\ref{sec:markov}. 

\begin{deluxetable}{llc ll}
\tabletypesize{\footnotesize}
\tablewidth{0.95\linewidth}

\tablecaption{\label{tab:phenomena}
State Change Phenomena in Pulsars
}
\tablecolumns{5}
\tablehead{
 \colhead{Phenomenon}
& \colhead{Properties} 
& \colhead{Number of }
& \colhead{Examples}
& \colhead{References}
\\
&& \colhead{States }
} 
\startdata
Nulls and bursts & On:off flux density ratio $\gtrsim 100$. 
			  & 2 to 4
			  & B0031$-$07
			  & 1,2,3
			  \\
                           & Null fractions: $\NF \le 94$\%.
                           & &  B1237+25 \\
                           
                           & Durations $\lesssim 1$~s to hours. & & B1944+17
                           &
                           \\
                           & Residence time PDFs: 
                           & &B0826-34 & 4, 5
                           \\ 
                           & \quad  1. broad, monotonic  
                           & & & 
                           \\ 
                           & \quad 2. dual components (often).
                           & & & 
                           \\
\hline                     
Mode changes & Changes in average pulse shapes. & 2 to 3 
		&  B0329+54   & 6,7 \\ 
		&&&  B1237+25 \\
\hline		
Subpulse drifts 	   & Systematic pulse-phase drifts & $\le 4$ 
				& B0031$-$07, & 8,9,10 \\
			   & \quad\quad through pulse window.         &             & B1822$-$09 \\
			   & Preferred drift rates. &&  B1944+17 \\
			   & Allowed, forbidden state sequences.&& B2319+60  \\
			   & Mixed with unorganized drift states. \\
			   & Quasi-periodicities  in 
			   $\gtrsim55$\% of pulsars. & & & 10, 11 \\
\hline
Long-term intermittency     & Off states: hours to years. & 2 &  J1832+0029 & 12  \\
	 &  Torque   1.5 to 2.5 times larger in on state.
 						&&   J1841$-$0500 & 13 \\
	& 38-day quasi-periodicity in B1931+24		&& B1931+24 & 14, 15 \\
\hline
Profile and torque changes & 	Correlated mode and torque changes. & $\gtrsim 2$ & & 16 \\ 
		& Month-like time scales. \\
		\hline
Radio/X-ray correlation &   X-ray on-off state transitions & 2 & B0943+10 & 17 \\
 & correlate with radio modes. \\ 
\hline
Extreme bursts &	Giant pulses.	& 2 &  Crab pulsar & 18 \\
                                                                   &&&(J0534+2200) & \\
                         &     Sporadic bursts with underlying period. & 2 &  Rotating radio &   19,20,21,22\\
                                                                                                     &&& transients & \\              
\enddata
\tablerefs{
1. \citet{1976MNRAS.176..249R};
2. \citet{1992ApJ...394..574B};
3. \citet{1986ApJ...300..540D};
4. \citet{2010MNRAS.408...40K};
5. \citet{2012MNRAS.424.1197G};
6. \citet{1982ApJ...258..776B};
7. \citet{2007MNRAS.377.1383W};
8. \citet{1970ApJ...162..727H};
9. \citet{1981A&A...101..356W};
10. \citet{1973ApJ...182..245B};
11. \citet{2006A&A...445..243W};
12. \citet{2012ApJ...758..141L};
13. \citet{2012ApJ...746...63C};
14. \citet{2006Sci...312..549K};
15. \citet{2013MNRAS.tmp..449Y};
16. \citet{2010Sci...329..408L};
17. \citet{h+13};
18. \citet{1995ApJ...453..433L};
19. \citet{2006Natur.439..817M};
20. \citet{2009ApJ...703.2259D};
21. \citet{2011MNRAS.415.3065K};
22. \citet{2011MNRAS.417.1871P}.
} 
\end{deluxetable}

\section{Markov and Stochastic Resonance Models}
\label{sec:markov}

The broad, exponential-like distributions for state changes, along with 
the lack of any correlation between adjacent null and burst durations, are naturally explained
by pulse sequences that conform to an underlying  Markov chain. So too is the inconsistency of nulls and bursts with the Wald-Wolfowitz `runs' test \citep{2009MNRAS.395.1529R}.  
An additional phenomenon --- stochastic resonance --- is suggested by the 
 binary state switching in   the long-term intermittent pulsar B1931+24.
The   durations of  on-and-off states for this object also have  wide distributions 
but they are neither  exponential nor monotonic.  
In addition, an underlying period of $38\pm5$ days is 
discernible in a 13-yr data set \citep[][]{2013MNRAS.tmp..449Y}.  
While sampling incompleteness may distort the 
distributions, the phenomena  are indicative of stochastic resonance in a nonlinear system
that has binary states and is driven by a forcing function that could be strictly periodic
or itself quasi-periodic.  

Along with binary states, some pulsars show multiple discrete values of subpulse drift rate or quasi-periods  in pulse amplitudes.   These have been compellingly  associated with a carousel  of sub-beams that rotate through the overall radio emission cone with circulation times of
tens to hundreds of seconds \citep[][]{1999ApJ...524.1008D, 2008AIPC..983..112R}.  The changes in drift rate or quasi-period, often combined
with nulling, imply that some  objects display at least four states.   These are also easily
describable as Markov chains but with some state transitions  disallowed, 
as discussed below.

Our approach is to generally discuss observed state-switching in terms of Markov processes
and stochastic resonance. 

\subsection{Markov Processes}

Markov processes are summarized here with reference to \citet[][]{Papoulis91}.
We consider $n$-state Markov processes (denoted  $M_n$) that have state values  
$\Svec = \{s_1, s_2, \cdots, s_n\}$.  Transitions between states are
described by an $n\times n$  stochastic matrix $\Qvec$ 
whose elements  are  the probabilities $q_{ij}, i,j = 1, \ldots, n$ for changing  from
state $s_i$ to state $s_j$ in a single time step. For observations we discuss,  the time step is a single rotation period.\footnote{Ultimately one must consider variations that occur on 
shorter time scales in the corotating frame and include the stroboscopic effect of pulse-window sampling related to rotation.   This kind of analysis is deferred to another paper.}
Normalization across a row is
$\sum_j q_{ij} = 1$ and  
the probability for staying in the $i^{th}$ state  is the
``metastability'' $q_{ii}$.  
The duration $T$ of the $i^{th}$ state is a random variable with
probability   $\propto q_{ii}^{T}$ for  $T$ time steps,  mean 
$\Tiave = (1-q_{ii})^{-1}$, and rms $\sigma_{T_i} = \Tiave \sqrt{q_{ii}}$.
The probability density function (PDF) of $T$  for the $i^{th}$ state is given by 
$q_{ii}^{T} = (1-T_I^{-1})^{T}$ normalized by the sum of this quantity over all values of $T$, or
\be
f_{T_i}(T) = 
\Tiave ^{-1}
\left( 1 -  \Tiave ^{-1} \right)^{T-1}, \quad T= 1, 2, \cdots.
\label{eq:pdf}
\ee
The PDF has constant slope  $d\ln f_{T_i}(T)/dT = \ln(1-1/T_i)$ and is generally steeper
 than an exponential that has the same mean, $T_i$.  However, as $T_i \to \infty$ (or equivalently $q_{ii}\to 1$), the PDF tends to a one-sided exponential function with $\sigma_{T_i} = \Tiave$. 
 
The ensemble probabilities of each state are given by the  state probability vector $\Pvec = (p_1, p_2, \cdots, p_n)$.    Stationarity, which we assume, requires that  $\Pvec\Qvec = \Pvec$,  so $\Pvec$ is the  left eigenvector of $\Qvec$ with unit  eigenvalue.   The transition matrix after  $t$ time steps, $\Qvec^t$,  converges  to a form $\Qvec^{\infty}$  whose rows are all equal  
to the state probability vector. The convergence time is some multiple of the longest-duration
state. 

A two-state $M_2$ process with $\Svec = \{n, b\}$ for nulls and bursts, for example, 
has $p_{1,2} = T_{1,2}/(T_1 + T_2)$. The nulling fraction  equals  $p_1$
and can be written as
\be
\NF = \frac{\Toneave}{\Toneave + \Ttwoave} = 
	\frac{1-q_{22}}{2 - q_{11} - q_{22}} = 
	\frac{q_{21}}{q_{12} + q_{21}},
\label{eq:nf1}
\ee 
making explicit that $\NF$ can be identical for objects that have markedly different 
mean state durations, as observed. 
Examples of  two-state Markov processes in 
Figure~\ref{fig:fig3} (top three rows) show the dependence
of the  time series and  state-duration histograms on the metastabilities
$q_{11}, q_{22}$.  

To model a time series, the number of states and the transition matrix need to be
determined. 
There are  $n(n-1)$ unique elements in $\Qvec$ for an $n$-state process,   
after accounting for normalization across rows.
Of these, $n$ elements can be determined from the mean state durations, 
$\Tiave, i=1,\dots,n$. Frequencies of occurrence of the states constrain an additional
$n-1$ elements. The remaining $n(n-3)+1$ elements need to be obtained using additional assumptions or measurements.   For $n=2$, the system can be determined solely
from the mean  state durations so measured frequencies of occurrence provide redundancy. For $n=3$, mean durations and frequencies need to be combined with one additional measurement is needed.    We will also consider $n=4$ processes that require five additional elements.  

\subsection{Nonlinear Systems and Stochastic Resonance}

Markov processes are often used to describe non-linear systems that have an
underlying potential-energy function with minima corresponding to the Markov states
\citep[e.g.][]{2012JChPh.136q4119B}.   Random 
motions (`noise') within potential wells induce state changes at rates that depend on well depths and noise strengths.    State changes can also be driven by a forcing function that
itself is stochastic, chaotic, or deterministic \citep[][]{a+06}.  A strictly periodic forcing function acting in concert with noise can induce state changes that are only quasi-periodic
\citep[][]{g+98}, a phenomenon known as stochastic resonance.

When driven by both noise and a forcing function $d(t)$, a system   potential having two local minima separated by a barrier can be described as an $M_2$ process 
in the adiabatic regime where $d(t)$ is slowly varying.  The   transition matrix is
\be
\Qvec = 
\left(
\begin{array}{cc} 
1-q_{12} e^{A_1 d(t)} & q_{12} e^{A_1 d(t)} \\
 q_{21} e^{-A_2 d(t)} & 1-q_{21} e^{-A_2 d(t)}
\end{array}
\right),
\label{eq:Qsr}
\ee
where $A_{1,2}\ge 0$ depend on properties of the wells and noise. 
Generally, the process is non-stationary. 
 For $A_{1,2} = 0$, $\Qvec$ reverts to the form for a simple $M_2$ process with
time independent elements.  Later we consider periodic forcing with
$d(t) = \cos(2\pi t/\Pf + \phi)$ where $\Pf$ is the period and $\phi$ is an arbitrary phase.
Over a cycle of $d(t)$, the well depths oscillate with the two wells out of phase by $\pi$.

\begin{figure}[h!]
\begin{center}
\includegraphics[scale=0.90, angle=0]{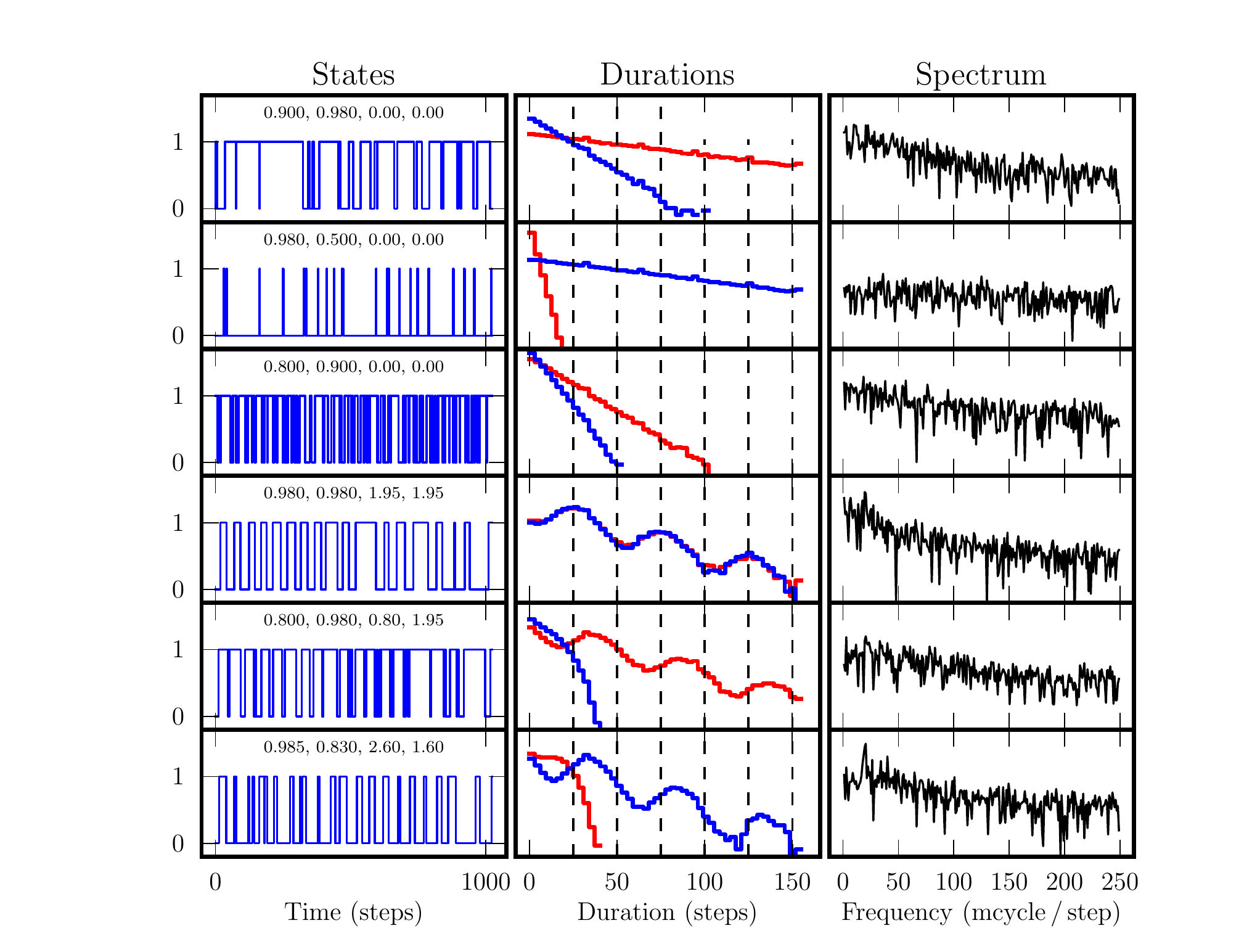}
\caption{Time series, histograms of state durations, and power spectra for two-state processes.
Left: single-realization time series for different values of the metastability probabilities
$q_{11}$ and $q_{22}$ and for the stochastic resonance amplitudes $A_{1,2}$, as given 
in the top of each frame.   
The top three rows are for pure Markov processes ($A_{1,2} = 0$) and the bottom
three include a periodic forcing function with $A_{1,2} > 0$ and period
$\Pf = 50$~time steps.
Center:  Histograms of the durations of  low states 
and high states 
states based on 5000 realizations of the time series.  The logarithmic vertical scales 
are from 1 to $10^{5.4}$. 
Right: Power spectra averaged over 5000 realizations plotted up to one-half the Nyquist
frequency. The logarithmic vertical scale
ranges from $10^{-7}$ to 0.1.  The 50-step period for the stochastic resonance in the lower three panels is evident
to varying degrees as a spectral line at a frequency of 20 milli-cycles~step$^{-1}$.
\label{fig:fig3}
}
\end{center}
\end{figure}

Examples of  two-state Markov processes in Figure~\ref{fig:fig3} with different degrees of stochastic resonance (bottom three rows) show how state-duration histograms depend on the system parameters. Unlike  the top three rows that show
exponential-like histograms, stochastic resonance makes the histograms
non-monotonic and periodic.  Power spectra shown in the right-hand panels are
featureless in the top three rows but show a spectral line to varying degrees in the bottom
three rows at a frequency $\sim 20$~mcycles~step$^{-1}$. 

\section{Application to Short-term Nulling and Pulse-Profile Mode Changes}

Nulls and bursts are often described empirically as two-state phenomena and so are some pulse-shape `mode' changes.   The null fraction appears to be characteristic of a pulsar
and is constant over decades in well-studied objects, suggesting that the transition matrix
in a Markov model is stable in the same manner as the average pulse profile.
Histograms of state durations are almost always  broad and monotonically
decreasing \citep[][]{1986ApJ...300..540D, 
2005MNRAS.357..859R,
2007A&A...462..257B,
2010MNRAS.408...40K,
2011ApJ...741...48C,
2012MNRAS.424.1197G} with many (e.g. B0835$-$41, B1112+50,  B1133+16, B2111+46, B2303+30) showing consistency  with the exponential-like form expected for an $M_2$ Markov process.   

Null and burst sequences fail the
 Wald-Wolfowitz `runs' test that is predicated on transitions occurring independently
 of the current state, as in coin tossing \citep[][]{2009MNRAS.395.1529R} or any
 independent, identically distributed (i.i.d.) process. 
 Coin tossing corresponds to a Markov process whose  
 transition matrix $\Qvec$  has
identical rows.   However, transition matrices consistent with observed nulls and bursts
necessarily have different rows and will fail the runs test.  The test statistic used for
the runs test is
$Z = (R-\langle R\rangle)/\sigma_R$ \citep[][]{2009MNRAS.395.1529R}, where R is the length of a null or burst; $Z$ has zero
mean and unit variance for an i.i.d. process. In 
 general an $M_2$ process has mean 
$\langle Z \rangle = \sqrt{N} (T_1^{-1} + T_2^{-1} - 1) = 
\sqrt{N}\left[(1-f_n) T_1^{-1} -1\right]$ where $N\gg1$ is the number of pulses analyzed. 
An i.i.d. process has $T_1^{-1} + T_2^{-1} \equiv 1$ so that $\langle Z\rangle = 0$. However
a Markov process  
with  nulling fraction
$f_n = 0.32$ and $N=1024$ pulses (an example in \citet[][]{2009MNRAS.395.1529R}) yields $\langle Z \rangle = -27.3$.   It is noteworthy that
estimates of $Z$ for 25 out of 26 pulsars range between -0.5 and -43 \citep[][]{2009MNRAS.395.1529R, 2012MNRAS.424.1197G}, as expected for an $M_2$ process that accounts for the observed state durations. 

\subsection{False Transitions from Measurement Noise and Beaming}

In this section we consider the effects of misidentified state transitions that inevitably
result from measurements with finite \snr.
Coherent emission from pulsars typically shows a broad pulse-amplitude distribution
\citep[e.g.][Figure 2]{2012MNRAS.424.1197G},
often taking a log-normal form \cite[e.g.][]{2012MNRAS.423.1351B} for the
phase-integrated intensity
 and in some
cases a power-law form 
\cite[e.g.][]{1995ApJ...453..433L, 1996ApJ...457L..81C, 2004ApJ...612..375C}.

A model for the   averaged intensity (pulse `energy')  in an on-pulse window is
\be
I(t) = b(t) A(t \vert M_2(t)) + N(t),
\label{eq:Imodel}
\ee
where $N(t)$ is white, Gaussian noise with variance 
$\sigma_N^2$ and $A(t\vert M_2(t))$ is a state-dependent 
pulse amplitude.   The time-dependent beaming function $b(t)$ is a multiplicative factor
that models pulse-to-pulse modulations  caused by the entire radiation beam
being only  partially and stochastically filled with 
sub-beams or containing  periodically spaced beams, as  in the carousel model
of \citet[][]{1999ApJ...524.1008D}.

Letting the two states be $\Svec = \left\{ 0, 1\right\}$, the pulse amplitude is
\be
A(t\vert M_2(t) = \left\{ 
          \begin{array}{ll}
	  	0 & M_2 = 0 \\
		a(\mu, \sigma) /\langle a \rangle & M_2 = 1,
		\end{array}
	\right.
\label{eq:Astate}
\ee
where $a(\mu, \sigma)$ is a log-normal random variable with parameters $\mu$
and $\sigma$. When $M_2 = 1$, a statistically independent value is drawn for
$a(\mu, \sigma)$ and normalized by the mean, $\langle a \rangle$.   Defined this way,  
$\langle A(t)\rangle = 1$ for the $M_2 = 1$ state
and the average signal to noise ratio is $\snr = \langle A \rangle / \sigma_N = 1/\sigma_N$.  
  
 \subsubsection{False Transitions from Additive Noise}
  
\begin{figure}[h!]
\begin{center}
\includegraphics[scale=0.70, angle=0]
{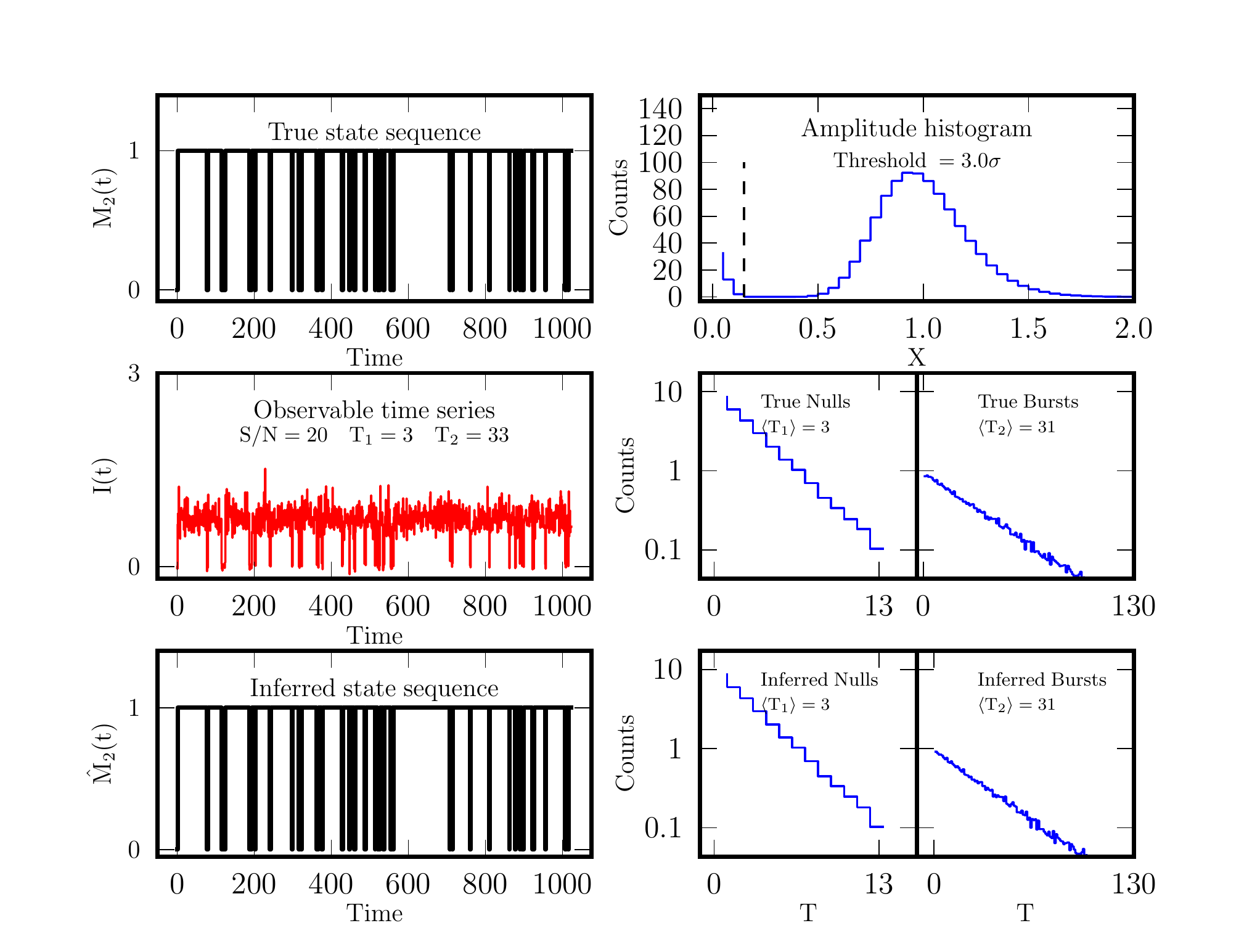}
\caption{Simulated time series and histograms for a case where all pulse amplitudes
are well above the threshold for dividing null and burst pulses.  The left column 
shows  a single  time-series realization while the histograms in the 
right-hand column are based on 1000 realizations.
Top left: time series of the Markov process.
Middle left:  time series of the intensity that
includes additive noise and pulse-amplitude variations (Equation~\ref{eq:Imodel});
the mean signal-to-noise ratio and mean state durations are indicated. 
Lower left: inferred state sequence obtained by identifying transitions in the intensity
time series.
Top right:  histogram of pulse amplitudes; the dashed line represents the $3\sigma$
threshold for separating null and burst pulses. 
Middle right: histograms of the true durations of null and bursts states; the estimated
mean durations are indicated. 
Bottom right: histograms of the null and burst durations inferred from the intensity. 
\label{fig:false_transitions1}
}
\end{center}
\end{figure}

\begin{figure}[h!]
\begin{center}
\includegraphics[scale=0.70, angle=0]
{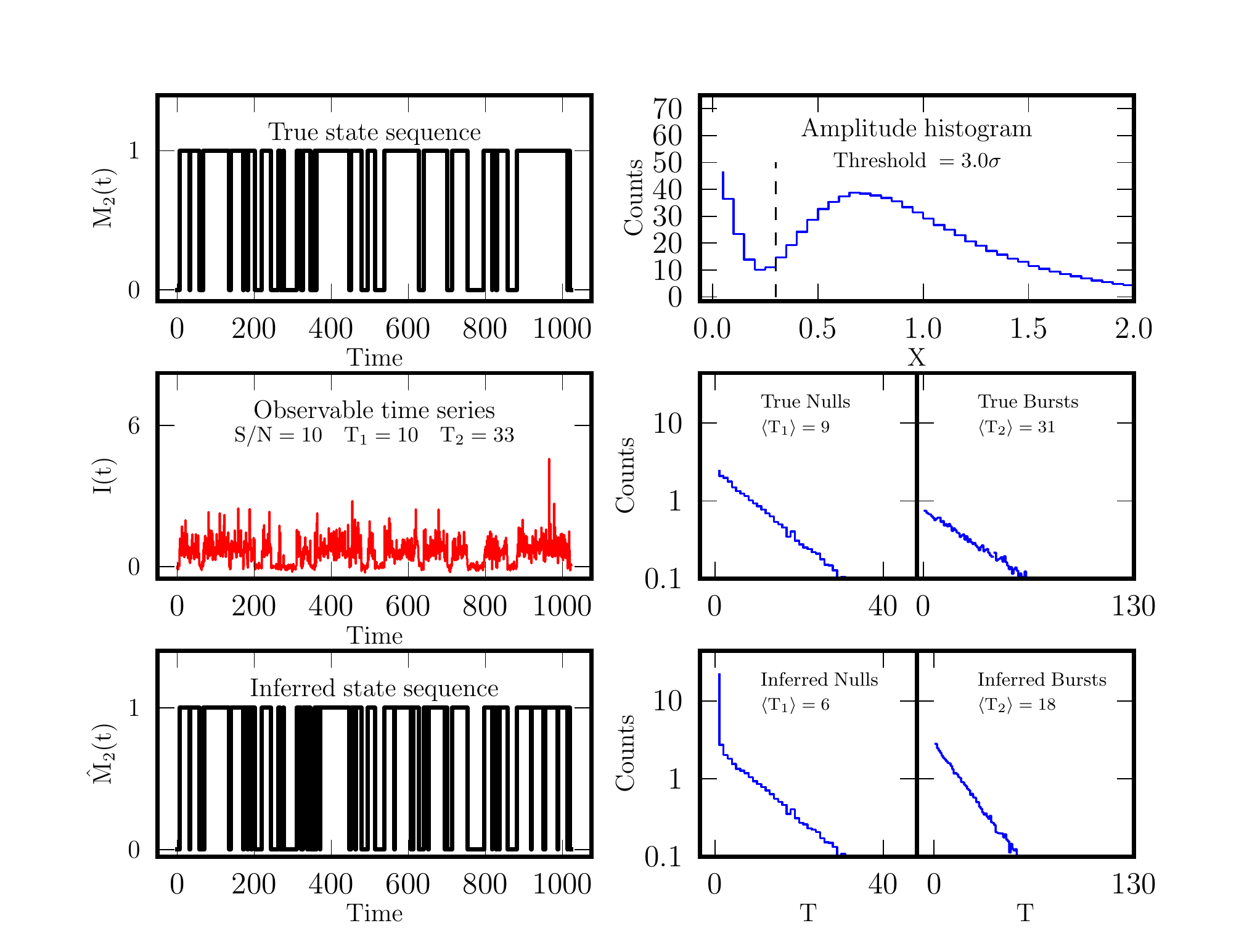}
\caption{Same as Figure~\ref{fig:false_transitions1} but for a case where the pulse
amplitude distribution and additive noise distributions overlap significantly. 
\label{fig:false_transitions2}
}
\end{center}
\end{figure}

First we consider the effects of additive noise when the beaming function is constant.
Figures~\ref{fig:false_transitions1} and \ref{fig:false_transitions2} show
time series and histograms for simulations using Equation~\ref{eq:Imodel}.
In Figure~\ref{fig:false_transitions1}  the low and high states
are well separated so that additive noise does not affect the identification of the
two states in a time series.  Time series were generated according to 
Equations~\ref{eq:Imodel}-\ref{eq:Astate}
using $\snr = 20$, $\mu = 1.5$, and $\sigma = 0.2$ and a threshold of 
$x=3\sigma_N$ was used to
discriminate between low and high states. 
The lower \snr\ case in Figure~\ref{fig:false_transitions2} with
$\snr = 10$, 
$\mu = 0.8$, and $\sigma=0.5$
shows overlap between the two observable state amplitudes, with consequent 
misidentified transitions.   These alter the duration histograms by adding many short
nulls and shortening the durations of bursts, which respectively cause the narrow feature
near $T=0$   in the inferred-null histogram and the steepening of the inferred burst
histogram in the lower-right of the figure.    

The false-transition probabilities for a threshold $x$ and for PDFs $f_N(N)$ and $f_a(a)$
for the noise and pulse amplitudes, respectively,  are
\be
r_{12} &=& \int_x^{\infty} dN\, f_N(N)
\\
\nonumber
r_{21} &=& \int_{0}^{\infty} da\,f_a(a) \int_{-\infty}^{x-a} dN\, f_N(N).
\ee
Given the statistical independence of  Markov transitions and noise-induced  threshold crossings, the combined  off-diagonal transition probabilities are (using `M' to denote the noise-free
Markov values)
\be
q_{12} = q_{12,M} + r_{12} - q_{12,M}r_{12} \ge q_{12,M},
\\
q_{21} = q_{21,M} + r_{21} - q_{21,M}r_{21} \ge q_{21,M},
\ee
showing that 
the interstate transition probabilities are always {\em increased} by additive noise,
as expected.

For the case presented in Figure~\ref{fig:false_transitions2}, $r_{21} > r_{12}$ because
the Gaussian noise PDF falls off more rapidly for high values than does the log-normal
PDF going to small values.   This causes the spike at short durations to appear only in the null-duration PDF.
For other choices of PDFs, \snr, and threshold, spikes can appear in one or the other (or both)
of the null and burst PDFs .

 \subsubsection{Pseudo-Nulls  from the Multiplicative Beaming Function}
 
 \citet[][hereafter KL10]{2010MNRAS.408...40K} identified an excess number of
 short nulls and short bursts over what would be expected from an $M_2$ process
 having long mean state durations. Their physical
explanation   is that the excess short  nulls are ``pseudo'' nulls resulting from  the granularity of conal carousel beams while true nulls (both short and  long) are caused by 
{\em bona fide} state changes, either a temporal cessation or a strong diminution of the radio
intensity.   This interpretation implies that the two processes (beaming and actual
nulling) are independent and therefore multiplicative. 

 When the beaming function $b(t)$ is included in the intensity model of
 Equation~\ref{eq:Imodel}  as a highly modulated quantity with
 minima well below the mean, the statistics of  true nulls and bursts are 
 modified. Short, pseudo nulls $\sim$one period in duration truncate bursts
 and therefore decrease the mean  durations of both nulls and bursts while increasing their numbers  in a  time series.    It can be shown that any modulation, periodic or otherwise, will have the same effect on the null and burst duration histograms.   It is not so clear that beam modulations will have the analogous effect of producing short bursts unless there are 
 occasional times when $b(t)$ is especially large for durations of one spin period (or less).

\subsection{Dual-Component Nulls and Bursts}
\label{sec:duals}

Several objects in addition to B1944+17 show dual characteristic durations  for nulls or bursts. 
Some but not all short nulls and bursts with $\sim 1$~period durations 
could be caused by false  transitions from additive noise or by pseudo-nulls due to 
multiplicative beaming variations, as discussed above.
 
 Here we acknowledge such contaminating effects on null and burst statistics but
 proceed with a Markov description, which is certainly allowed no matter the cause
 of the various kinds of nulls and bursts. 
  In particular, we demonstrate  that the estimated distributions
for null and burst durations can be accounted for by an $M_4$ process.   
The state set  $\Svec = \{n_1, n_2, b_1, b_2\}$ comprises
 two null states ($n_{1,2}$) and  two burst
 states ($b_{1,2}$) that are not observationally distinguished, apart from 
 their durations.\footnote{Models that include multiple states that are indistinguishable observationally are hidden Markov models 
 \citep[HMM;][]{r89} in which the underlying physical states are manifested indirectly.
 We note that such models have been used to identify state changes in a pulsar
 \citep[][]{2011MNRAS.415..251K}}
 To proceed, we adopt a  minimalist model where transitions 
 are allowed only between a
 null state and a burst state and {\em vice versa} but $n_1$ - $n_2$ and $b_1$-$b_2$ 
 transitions in either direction are forbidden. 
 Of the 16 equations needed to solve for the matrix elements, four come from
 normalization across rows, four from the mean state durations, and three from
 the frequencies of occurrence of the four states, which we estimate from information
 in KL10.
  The forbidden transitions are represented by four zero elements, so only one additional
 equation is needed.   
 We estimated mean durations 
 $(T_1, T_2, T_3, T_4) \approx  (1.01,  16.6,  1.01, 8.9)$ (in period units)
 and frequencies of occurrence for the four states, 
 $\fvec = (f_1, f_2, f_3, f_4) \approx (0.01, 0.655, 0.013, 0.321)$, from Figure~3 
 of KL10. 
 
 The full transition matrix was determined by searching over
 a grid of values for the off-diagonal elements,
 q13, q23, q31, and q41, while holding the diagonal elements $q_{ii}, i=1,4$ fixed
 to values consistent with the mean durations
 and setting $q_{12} = q_{21} = q_{34} = q_{43} = 0$.  For each trial transition matrix
 $\Qvec$ we determined the state probability vector $\Pvec$ from $\Qvec^{100}$ (as
 an approximation to $\Qvec^{\infty}$; see Section~\ref{sec:markov}). 
 By minimizing the mean-square difference $\vert \Pvec-\fvec\vert^2$
 we obtained a (non-unique) solution,
  \be
\Qvec_{(n_1,n_2,b_1, b_2)} \approx \left( \begin{array}{cccc}
0.01 		& 	0 		& 	0.886 & 	0.104 \\
0      		&	0.940 	&	0.006	&	0.054 \\
0.469	&	0.521	&	0.01		& 0 \\
0.012	& 	0.100	&	0		& 0.887
\end{array} \right).
\label{eq:4state}
\ee
The values of 0.01 for the metastabilities of short nulls and short bursts  yield
durations of order one time step, i.e. $T = 1/(1-0.01) \sim 1$. 
Higher time resolution data can sharpen these values.  Figure~\ref{fig:fig2} (left) shows
the Markov distribution that superposes the two burst states (properly weighted
according to their frequencies of occurrence and durations) overplotted on 
the histogram of burst durations from KL10.    The Markov distribution is a good representation
of the salient features of the histogram.   There may be an excess of observed long bursts 
compared to the observed counts, but they are small in number;
 some of these may result from  short nulls that are missed in  some
of the data (i.e. false {\em non}-transitions). 
 
 As implied in our merging of the two burst states,  
 the four-state description for B1944+17 can be reduced to a three-state description if
 short and long nulls are combined into an 
 $\Svec = \{n, b_1, b_2\}$ model or bursts combined into an  $\Svec = \{n_1, n_2, b\}$
 model.  
Similar 3-state descriptions can account for the dual nulling patterns seen from
pulsars 
B0809+74 \citep[][]{2002A&A...387..169V, 2012MNRAS.424.1197G};
B0818$-$13 \citep[][]{2004A&A...425..255J};
J0828$-$3417, J0941$-$39, and J1107$-$5907 \citep[][]{2012MNRAS.423.1351B};
B0835$-$41,  B2034+19, B2021+51, and B2319+60 \citep[][]{2012MNRAS.424.1197G}.
The object B0826$-$34 shows a  null state with occasional single pulses and a burst
state with rare single-pulse nulls \citep[][]{2012ApJ...759L...3E} for which
the average pulse shapes in these states differ significantly.   This object is also 
amenable to a Markov description but a quantitative description of state durations
and fractions is needed.

 \subsection{Subpulse Drift Modes}
 
 In some pulsars, there are well-organized motions 
  of subpulses across the pulse-phase window defined by averages of a large number of pulses.    Two or three subpulses  typically appear  in 
 a single pulse period with separation $P_2$, usually expressed in time units.  In a sequence of
 pulse periods, bands  of drifting subpulses have separations $P_3$ at fixed
 pulse phase, usually
 expressed in units of the spin period $P$.  The drift rate is then $P_2/P_3$ or, in
 physical units, $\dot\phi = P_2/P^2P_3$~cycles~s$^{-1}$ when $P_2$ and 
 $P$ are expressed in time units and $P_3$ in spin periods.  
 A common interpretation is that drifts correspond
 to $\Evec \times \Bvec$ motions of multiple particle and radiation beams
 (a beam `carousel') around a magnetic axis or some other related axis
 \citep[][]{1975ApJ...196...51R, 1999ApJ...524.1008D, 2006ApJ...650.1048G, 2012ApJ...746...60L}.
 The total time needed for a beam to circulate around the axis is
 $P_4 = 1/\dot\phi P = PP_3/P_2$ in units of  spin periods\footnote{$P_4$ is sometimes
 denoted $\hat P_3$ \citep[e.g.][]{1975ApJ...196...51R, 2008AIPC..983..112R}, which can
 be confused with $P_3$ and is therefore not used here.}.  While $P_2$ and $P_3$  are
 visually recognizable in pulse sequences for some pulsars, the circulation time $P_4$ is often
 inferred from power spectra of pulse intensities at a single pulse phase
 \citep[][]{1973ApJ...182..245B, 1981A&A...101..356W, 1999ApJ...524.1008D, 2006A&A...445..243W}.
 For some objects discussed below, the drift pattern can have variable
  $P_3$ and $P_4$ but with near constancy of $P_2$.

 When in the `on' state, B1944+17 shows additional complexity in the form of 
four different  emission modes  \citep[][KL10]{1986ApJ...300..540D}
that are distinguished by different subpulse drift rates or average pulse shape.   Combined with nulling described above, as many as ten states may be needed for a 
full description, but additional observational classification is needed for any detailed modeling.  

Two pulsars that
show three distinct drift rates along with nulls are
B0031$-$07 \citep[][]{1970ApJ...162..727H,1981IAUS...95..211W,1997ApJ...477..431V}   and B2319+60 \citep[][]{1981A&A...101..356W}.  The  states are labelled  A, B, and C
in order of increasing drift rate\footnote{For B2319+60,  we denote the
`abnormal' (ABN) mode defined by \citet[][]{1981A&A...101..356W} as `C'}.
For B0031$-$07,  the drift rates have ratios
 1:1.9:2.8 and observed sequences include B only and hybrid AB and BC bursts.
There are no A only or C only bursts, or AC, BA, CA, or CB combinations.   
A four-state model $\Svec = \{n, A, B, C\}$ will have a transition matrix with
$q_{nC} = q_{An} = q_{AC} = q_{BA} = q_{CA} = q_{CB} = 0$ and  six  remaining 
unique values need to be determined from measurements.  Three can be obtained from the
frequencies of occurrence $\fvec = (f_n, f_A, f_B, f_C) \approx (0.45, 0.086, 0.45, 0.014)$
and the four diagonal elements from the as-yet unknown mean durations of each state. 
Future measurements can allow determination of the full matrix with one redundant estimate
that can be used as a check.   Our discussion comprises an existence proof for a Markov model because, even without knowing the diagonal elements of $\Qvec$, it provides the mechanism for  `forbidden' transitions as well as the allowed transitions.

Very similarly, B2319+60 has A,B, and C states that display different drift rates and 
pulse shapes along with nulls occurring with 
$\fvec = (f_n, f_A, f_B, f_C) \approx (0.3, 0.45, 0.15, 0.1)$. However,  there are more allowed sequences than for B0031$-$07, so eight unique elements
of the transition matrix need to be determined
after setting $q_{AC} = q_{BA} = q_{CA} = q_{CB} = 0$. Seven can be estimated from the mean state durations (currently not in the literature) and $\fvec$.  The eighth can be obtained by
estimating the number of AB or BC transitions.  

A remarkable  feature of state transitions in both B0031$-$07 and B2319+60
is that   subpulse drift states  occur only in order of {\em increasing} drift rate.  
In the carousel model, the 
drift rate is related to the local plasma drift velocity $c\Evec\times\Bvec/B^2$ that vanishes
when the local charge density equals the Goldreich-Julian density.   An increasing drift rate
therefore suggests depletion of the charge density over the course of a burst that 
eventually terminates
in a null.  In a nonlinear model, the systems in these two pulsars  seem to be 
running `downhill' in an effective 
potential that is somehow reset during and perhaps because of a null.  The A, B, and C
states would have energy minima such that it is much more likely to
have AB, BC, and AC  transitions than BA, CB, or CA transitions. 

\subsection{Mode Changes with Long Time Constants}

Two pulsars show pulse-shape modes that have finite time constants for the evolution
of the drift rate in one of the modes.  The exemplar drifting-subpulse object
B0809+74 has 1.4\% null pulses that interrupt bursts 
\citep[][]{1983MNRAS.204..519L,2002A&A...387..169V}. 
Like other objects, the durations
of nulls and bursts have Markov distributions, but during bursts the drift rate
{\em decreases} at the onset of a null and relaxes  exponentially to a larger
 asymptotic value over a time that scales with the null duration but is typically tens of seconds
\citep[][]{1983MNRAS.204..519L}.

The B and Q burst modes from B0943+10 
\citep[][]{2008AIPC..983..112R}
can be accommodated by a two-state
Markov process but with some complications. 
This pulsar displays  correlated X-ray and radio states that, along with torque variations seen in 
other pulsars, link radio and high-energy phenomena \citep[][]{h+13}.  
The two burst-only radio states  show a quasi-period $P_4$ that is constant in
the weaker Q mode but evolves exponentially with a 73~min time constant in the B mode.
As $P_4$ increases the pulse shape changes progressively on the same time scale. 
The exponential increase  
in $P_4$ (equivalent to a decrease in  drift rate)  suggests in the carousel model that,
unlike for B0031$-$07 and B2319+60 discussed above,  the 
$\Evec \times \Bvec$ drift velocity decreases in B-mode sequences, implying
that the polar-cap acceleration region trends toward a force-free state.   In the ionic evolution model of \citet[][]{2011MNRAS.414..759J},
the vertical potential drop in the polar-cap accelerator evolves as high-Z ions are increasingly
 ionized and disassociated.    It is not clear in either of these interpretations why the state changes are not more periodic rather than showing a wide range of durations, but a 
nonlinear system driven stochastically  can account for these.

\section{Long-term Intermittency and Stochastic Resonance}
\label{sec:SR}

Three objects (J1832+0029, J1841-0500, and B1931+24) show distinct on and off 
emission states  
that are accompanied by switching between two values of the spindown rate $\dot\nu$.   
Thus far there are no detailed descriptions of pulse variations in the on state because
intensity and timing results have been based on  
  average profiles rather than single pulses.   
  
  The best studied object, B1931+24,
  shows no  short-term nulls (minutes or less) when the pulsar is in the
 long-term `on' state \citep[][]{2006Sci...312..549K}.   The durations of ons and offs
 have mean values of $8\pm4$~d  and $22\pm7$~d and large maximum to minimum
 ratios,  19:1 and 10:1, respectively \citep[][]{2013MNRAS.tmp..449Y}.  
 These ratios  are similar to
 those of short-term nulls and bursts for other objects and, along with 
 the lack of correlation between the durations of 
 contiguous on-state and off-states,  are consistent with a Markov  description.

 \begin{figure}[h!]
\begin{center}
\includegraphics[scale=0.4, angle=0]
{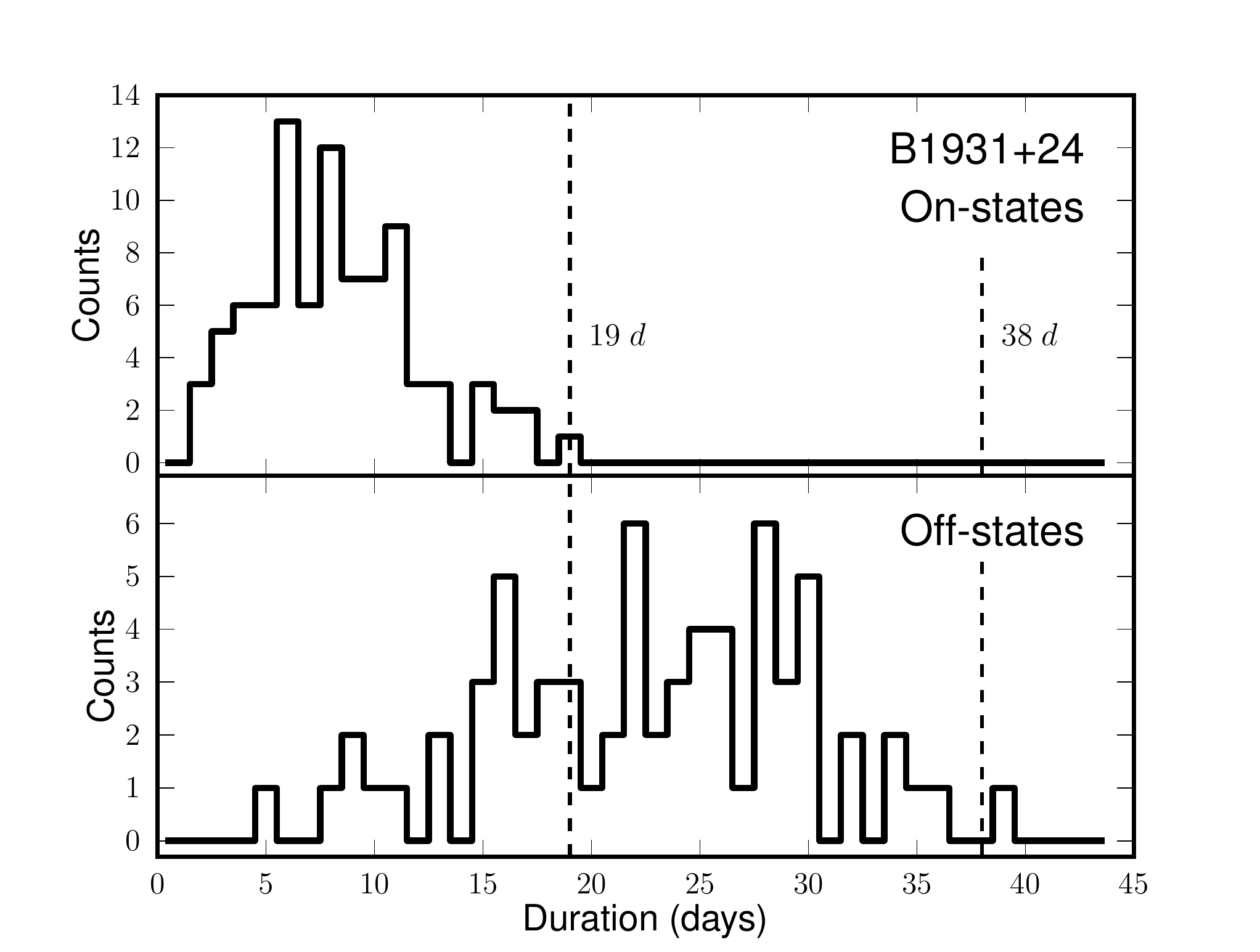}
\includegraphics[scale=0.4, angle=0]
{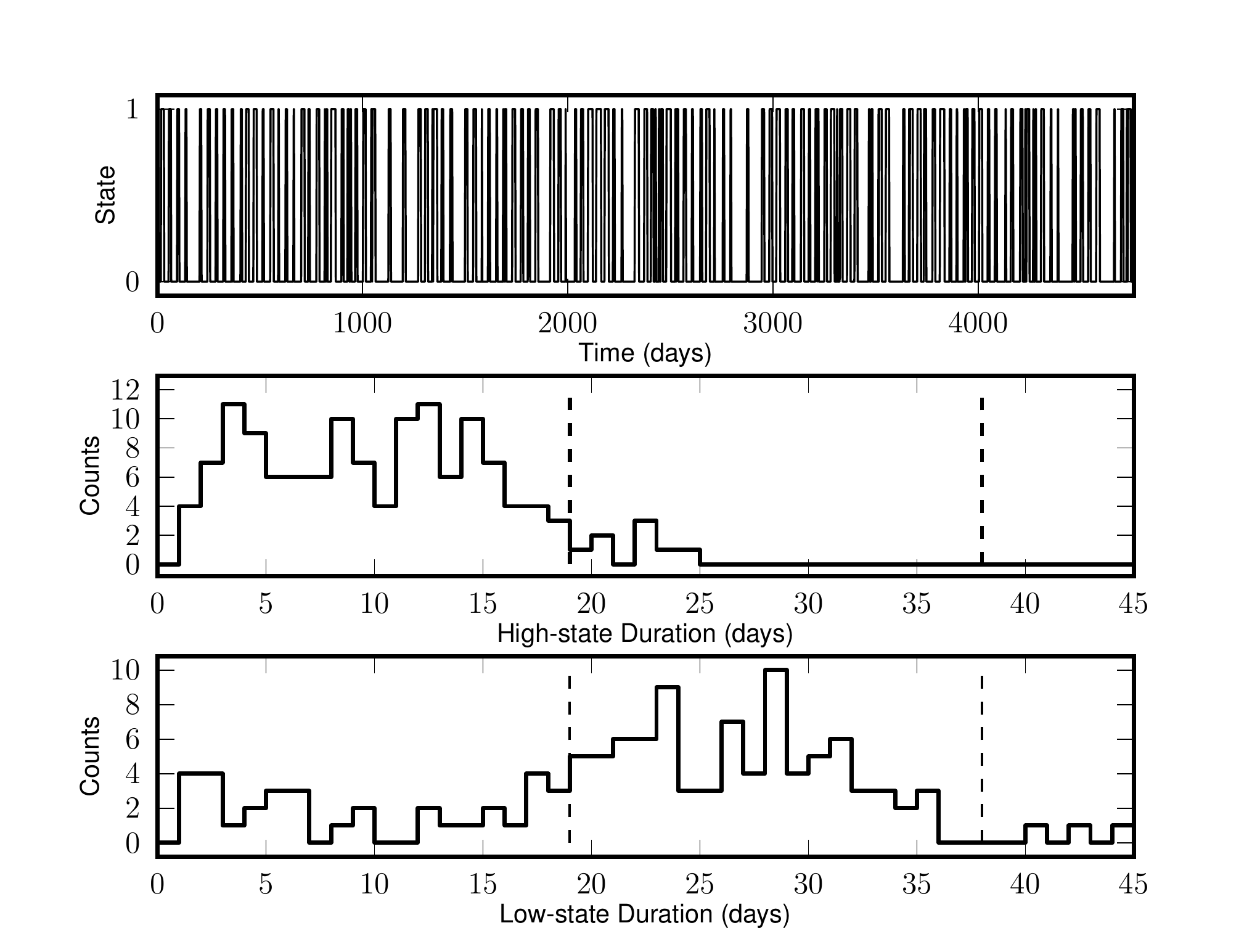}
\caption{Left: Histograms of on-and-off state durations for B1931+24, after Figure~2 of
\citet[][]{2013MNRAS.tmp..449Y}. Vertical dashed lines indicate one-half and a full period
of $P_i = 38\pm4$~d identified in a 13~y data set \citep[][]{2013MNRAS.tmp..449Y}.
Right: Time series of state changes and histograms or state durations for a stochastic
resonance model of the form of Equation~\ref{eq:Qsr} with
$q_{11} = 0.985, q_{22} = 0.83, A_1 = 2.6, A_2 = 1.6$, and $P_f = 38$~d. 
\label{fig:B1931hists}
}
\end{center}
\end{figure}

 However, the histograms of state durations  are inconsistent with the monotonic form expected
 from a Markov process.   
 As shown in Figure~\ref{fig:B1931hists} (left panel), based on Figure~2 of 
 \citet[][]{2013MNRAS.tmp..449Y} 
 (see also Figure~1c of \citet[][]{2006Sci...312..549K}), 
 the mean state durations are  about equal to their modes, and there is
 a paucity of short durations. 
 The overall small number of on-off cycles in the 13-yr data
 set combined with sampling incompleteness   (from allowing      
 observation gaps as large as 5~d  in the analysis, 
 \citet[][]{2013MNRAS.tmp..449Y}) may contribute to the on-state histogram shape.  
Conversely, the deficit of off-state   durations  up to   $\sim 15$~d  is not plausibly accounted for this way.

Another difference from fast intermittency and Markov behavior is   an apparent
underlying   period  for the intermittency, $ P_{\rm i} = 38\pm5$~d, in the 13~yr data span \citep[][]{2013MNRAS.tmp..449Y} that is  longer
than any of the individual on-and-off  durations and is about 4.8 times the mean on-duration
and 1.7 times the mean off-duration.   The quasi-periodicity may be interpreted in two ways.     First, there may be a  
strictly periodic --- but hidden --- process that is blurred by mediating processes that
lead to observable quantities.   Alternatively,   the 38~d period may  be the mean of a  process that is intrinsically quasi-periodic.    We explore the first of these options because it
is  more specific.  

A stochastic-resonance model is an acceptable statistical description of
observations of   B1931+24
if we adopt a transition matrix  (Equation~\ref{eq:Qsr}) with a sinusoidal forcing
function $d(t) = \sin(  2\pi t/ P_f +\phi)$. 
In this interpretation
the different mean on-and-off durations $T_{\rm on}, T_{\rm off}$ signify an asymmetric effective potential and their sum,    
 $T_{\rm on} + T_{\rm off} = 31\pm 8$~d,  is nominally smaller than  the observed mean period $P_i = 38\pm5$~d reported by \citet[][]{2013MNRAS.tmp..449Y}.     Schmitt triggers 
  show similar behavior when the   underlying potential has asymmetric minima
  \citep[][]{m+99}.

The right-hand  panel of Figure~\ref{fig:B1931hists} shows state-duration histograms for
a stochastic-resonance model that is consistent with the observed histograms in 
the left-hand panel.   The simulations include a strictly periodic forcing function with
 $P_f = P_i$. The centroids of the dominant features in the simulated histograms 
 straddle the half period $P_f/2 = 19$~d, as is common in SR models.  There is an additional
feature in the simulated off-state histogram for durations $\lesssim 7$~d that may be absent
in the actual data because of the incomplete sampling of short-duration states 
\citep[][]{2013MNRAS.tmp..449Y} and 
because the total number of counts is smaller than in  the simulated histogram. 
A more precise test of stochastic resonance in  
 B1931+24 would require longer, continuous data sets than are currently
available.   Data sets that include single-pulses would be valuable for characterizing
any very short-duration on or off states. 
 
The other two objects that show long-term intermittency and discrete values
of $\dot\nu$, J1832+0029 and J1841-0500, have state durations that are longer
than those in B1931+24 and have not been sampled adequately  to test whether 
a stochastic resonance model might apply to them as well.

\section{Timing Variations from a Two-State Markov Process}
\label{sec:timing}

Objects that are intermittent on long time scales
 like B1931+24 show a distinct value of spindown rate 
$\dot \nu$ in each state.   
Switching between discrete values of $\dot\nu$ can be viewed as  
  pulse-like changes in $\dot\nu$ (i.e. equal upward and downward transitions)
  that combine with 
 a constant value of $\dot\nu$, where we assume that the second derivative
 $\ddot\nu$ from the spindown torque is negligible.
Viewed collectively as a  
two-state Markov process, switching  between  spindown rates produces  a random walk in spin frequency and an integrated random walk for pulse phase. 
Stochastic timing variations occur  
because individual states last for times that vary widely about their means,
 as discussed earlier.
 We integrate  the fluctuating
 part of $\dot\nu(t)$ 
 twice to get the spin phase perturbation and calculate its variance over
 a data span of length $T_d \gg \Tbar$ .  Using the `reduced' duration,
 $\Tbar = T_1 T_2 / (T_1 + T_2)$, the
  rms timing variation expressed in time units can be written as   
 \be
 \sigma_{t,2}(T_d)  =  C_2^{-1} \sigma_t(T_d)
 	\approx 74~\mu s \,P \,\vert\dot\nu_{-15}\vert\, \Tbar_3^{1/2} T_{d, yr}^{3/2} 
	\left\vert\frac{\delta\dot\nu}{\dot\nu}\right\vert
	\frac{\sqrt{f_l(1-f_l)}}{1/2}.
\label{eq:rmstiming2}
\ee
In this equation, we have used $\Tbar = 10^3 \Tbar_3$~s, $T_d$ is in years, and
$\dot\nu = \dot\nu_{-15} 10^{-15}$~s$^{-2}$.  
The factor
$C_2 \approx 15.5$ corrects for the fluctuations removed by subtraction 
of the second-order polynomial  fit for the mean spin rate and $\dot\nu$. 
All else being equal, 
 the rms timing error is a maximum for equal fractions of low and high states,
 $f_l = f_h = 1-f_l = 1/2$.   The  scaling, $\sigma_t(T_d) \propto T_d^{3/2}$,  
 is generic to random walks in spin frequency
 \citep[][]{1975ApJS...29..443G, 1985ApJS...59..343C, 2010ApJ...725.1607S}. 

For B1931+24,  individual transitions can be identified and removed from the data
to reduce the rms timing variation \citep[][]{2006Sci...312..549K}.
However for other objects, individual states may not be identifiable because they are too small
or are mixed with other spin variations and consequently will appear as spin noise
with nonstationary variance. 

\section{Medium-term Intermittency and Timing Variations in B0823$+$26}
\label{sec:B0823+26}

B0823+26 ($P = 0.53$~s and $\dot\nu = 6\times10^{-15}$~s$^{-2}$)  displays 
 minute-duration nulls along with those of hours and longer.
Any emission during nulls is less  than 1\% of  the mean on-state flux density.
The distribution of on-state durations shows both a narrow and a broad component with
times scales of minutes and days, respectively \citep[][]{2012MNRAS.427..114Y}. 
The durations of  particular nulls and bursts may be mis-estimated in some cases due
to data windowing effects that cause state changes to be missed, as noted by \citet[][]{2012MNRAS.427..114Y}, but the presence of both  long and short nulls
is secure.

 Torque variations  are evidently more complex in this object than in 
 long-term intermittent objects like
 B1931+24. In a 153-d data set, \citet[][]{2012MNRAS.427..114Y} saw neither cubic
 structure in the timing residuals above an rms residual of about  200$~\mu s$ nor 
  switching between  discrete values of $\dot\nu$ like that seen in B1931+24.  
  The upper bound on the amplitudes of such pulses
  is $\Delta\dot\nu / \dot\nu\lesssim 0.06$ \citep[][]{2012MNRAS.427..114Y} based
  on detailed modeling of the time series.
  What {\em are} seen in a longer data set,
  however, are step-function-like changes in $\dot\nu$  (rather than pulses) with   amplitudes
 as large as $\vert \Delta\dot\nu / \dot\nu \vert = 0.02$ but occurring 
  at a rate of only $\sim 1.2$~yr$^{-1}$
  \citep[][]{1985ApJS...59..343C}, much less frequent  than the state changes discussed here.
  
  We can place an additional statistical limit on the amplitudes of pulses in $\dot\nu$.  
Equation~\ref{eq:rmstiming2} with $p_1 = 0.2$,
 $T_1 = 0.26$~d and $T_2 = 1.4$~d \citep[][]{2012MNRAS.427..114Y} predicts
 $
\sigma_{t,2}(153~\rm d) \approx 610~\mu s~ \vert\delta\dot\nu/\dot\nu\vert.
$
  Comparison with  
  the upper bound of 80~$\mu s$ on the rms
  residual (c.f. Figure~8 and discussion in \citet[][]{2012MNRAS.427..114Y})
  implies 
$\vert\delta\dot\nu / \dot\nu\vert \lesssim 0.13$, a less stringent 
constraint than $\vert\delta\dot\nu/\dot\nu\vert \lesssim 0.06$ obtained by
\citep[][]{2012MNRAS.427..114Y} from direct fitting.   Using the smaller upper bound,
we estimate that for a 13.6-year data set like that analyzed by \citet[][]{1985ApJS...59..343C}, state
changes would produce no more than 
$\sigma_{t,2}(13.6~\rm yr) \approx 6.7~ms$,
much smaller than the 
 measured value of $\sim 100$~ms over this data span.   Spin noise in B0823+26 is
 evidently 
dominated by step functions  in $\dot\nu$ that yield a scaling
$\sigma_t(T_d) \propto T_d^{5/2}$, much steeper than that expected from pulses in
$\dot\nu$.   The origins of the step functions in $\dot\nu$  for B0823+26 appear distinct from the process that causes large pulses in $\dot\nu$ for B1931+24 \citep[][]{2006Sci...312..549K}.

From the timing analysis on B0823+26 we conclude the following.  First, switching between
nulls and bursts on time scales
of days evidently is not associated with large jumps between binary values of $\dot\nu$, the same conclusion as reached by \citet[][]{2012MNRAS.427..114Y}.
Second, the large amount of spin noise appears to have other
causes than those that are responsible for the intensity jumps.   The difficulty in identifying
discrete values for $\dot\nu$ in B0823+26 may result from a two-state model being too
simplistic.  Any   correlated torque events might also have different shapes than pulse-like forms. If the torque does not return to its original value (as it does in B1931+24 and other
intermittent pulsars), the spin noise will appear more like that observed.    Multiple
states are also implied by the presence of  nulls and bursts that each
have short and long-duration components.  
A timing analysis of multiple state models  is beyond
the scope of this paper but  is worthy of future analysis.

\section{The Physics of Discrete States and Triggering Between States}
\label{sec:states}

The aggregated observational results presented in this paper demonstrate that diverse
phenomena  result from the existence of metastable states in  the dynamics of pulsar magnetospheres.  A physical understanding of the phenomena leads to two primary questions: 
(1) What is the underlying magnetospheric physics of the observed states?
and 
(2) What causes transitions between states?
We discuss these questions in the context of standard ideas about pulsar magnetospheres.


Some characteristic light-travel times associated with neutron stars (NSs) and 
their magnetospheres are useful for the following discussion. 
The NS radius  is  $R_* \sim 30~\mu s$ and the radius of the  PC is
 $\rpc = R_* (R_* / \rlc)^{1/2}\sim 0.5P^{-1/2}~\mu s$ for spin periods $P$ in seconds, where 
the light cylinder radius $\rlc = cP/2\pi \sim 0.16P~s$.  In models with an acceleration region
just above the PC, the height of the region $h \lesssim \rpc$ and for objects with 
sustained pair cascades,  $h \sim 0.3~\mu s$.   Radio intensity variations are seen on 
these and smaller time scales.   Indeed, {\em a priori} one might expect {\em all} observed
time scales to be no larger.   However, the variations discussed in this paper concern
macroscopic scales that exceed the characteristic times  by many orders of magnitude 
(seconds to years) but are still much smaller than the spindown
time $\tau_s = P / \dot P \sim  10^{7.2}P \dot P_{-15}^{-1}$~yr on which the magnetosphere 
evolves (where $\dot P = 10^{-15} \dot P_{-15}$~s~s$^{-1}$ is the period derivative).
Thermal time scales  enter the picture when thermionic emission plays
a role and the temperature of  a hot PC is time variable.   Also, in some situations,
pair cascades are temperature dependent \citep[][]{2001ApJ...554..624H, 2007MNRAS.382.1833M}.  The heating and cooling
times of the PC are estimated to be $\lesssim 1$~s and thermostatic regulation 
from pair production may result in variations on some multiple of this time scale \citep[][]{1980ApJ...235..576C}.
Surface physics introduces 
 the proton drift time through the atmosphere $\tau_p \sim 0.1$-1~s
and the `excavation' time for removing
one radiation length's amount of material from the PC,
$\tau_{\rm rl} \approx 2.1\times 10^5~s~(P/Z B_{12})$ \citep[][]{2011MNRAS.414..759J, 2012MNRAS.423.3502J}.  Orbital periods around a 1.4~$\Msun$ NS are
$\Porb \approx 127~s\,P^{3/2}(r/\rlc)^{3/2}$ using $\rlc$ as a fiducial radius.   
An object with a 38~d orbital period at  $r\sim 0.3$~AU is well outside
the gravitational  tidal disruption radius, 
$r_{tg} = (3M_{\rm NS} / 2\pi\rho)^{1/3} \sim 10^{11}\rho^{-1/3}$~cm.
Small objects inside the tidal radius can exist owing to tensile forces.

Radio emission requires relativistically beamed, coherent radiation from a particle flow that is collimated in a narrow, magnetic field-line bundle.   Coherence is produced by an
instability that requires counterstreaming particle flows and has a maximum growth rate at an altitude-dependent plasma frequency.   Also required is a favorable viewing geometry that
depends on the angular width of the radiation beam, which varies with altitude, and on
the angle between the beam and the observer's direction.   
The observed discrete states therefore require changes in one or more 
of these elements  to alter the intensity, pulse shape, and spectrum
of the radiation.     That discrete states  are  seen in the
dominant energy loss channels --- viz. spindown rates $\dot\nu$ in the long-term intermittent
pulsars like B1931+24 and X-ray emission from B0943+10 --- implies that  significant, 
discrete changes occur in the  magnetosphere at large.  

\subsection{Metastable States}

Several  gross features of pulsar magnetospheres could provide binary states, 
including on-and-off modulation of  $e^{\pm}$ production, cycling of the relative contributions of ions and pairs to the current flow, or modulations of the total current itself between two extremes, such as  a vacuum state (no current) and  a force-free state 
\citep[e.g.][]{2012ApJ...746...60L, 2012ApJ...746L..24L}. 
Another dyadic pair could be  a dead, `electrosphere' state 
\citep[e.g.][]{1980Ap&SS..72..175M, 1985A&A...144...72K, 2001MNRAS.322..209S, 2002A&A...384..414P}~{\em vs.} a standard, dynamic magnetosphere. 
Extreme switching between an electrosphere and a magnetosphere  might be associated
with the more extreme RRATs and long-term intermittent objects
\citep[e.g.][]{2010HEAD...11.1621M}.

Observed state changes so far are exclusively the realm of older, canonical pulsars with
periods $\gtrsim 0.3$~s and magnetic fields $B\sim 10^{12}$~G.   Indeed, there appears to be 
a greater propensity for state changes in pulsars that are longward of $\sim 0.3$~s in the $P$-$\dot P$ diagram \citep[Figure 1,][]{2008ApJ...682.1152C}.
However, it is conceivable that shorter period objects display state changes that have been missed because of  observational selection, particularly MSPs for which there are few observations of single pulses.   Whether any state changes are {\em expected} from MSPs
depends on the nature of the states themselves and on how state changes are triggered. 
If one of the state dyads is associated with a threshold for pair production, state transitions
will be expected if the pair-production death line is near the MSP population in 
the $P$-$\dot P$ diagram.  Published death lines do not provide clarity on this because 
they depend on the types of current flow (SCL or not), on the photon-emission process
(curvature vs. inverse-Compton) 
\citep[][]{2001ApJ...554..624H, 2002ApJ...576..366H, 2011ApJ...736..127W, 2011ApJ...743..181H, 2007MNRAS.382.1833M}, and on any offset of
PCs from the dipole axis caused by current-driven sweepback of the magnetic field
\citep[][]{2011ApJ...743..181H} or by an offset dipole \citep[e.g.][]{2001ApJ...554..624H}.   If alternative photon-processes can drive pair cascades
for different regions in the $P$-$\dot P$ diagram, the propensity for state changes may
be associated with just one of the processes, such as non-resonant inverse Compton
radiation vs. curvature radiation.

\subsection{State Triggering}

A model is proposed in which the dynamics of  current flows are responsible for both defining discrete states and for causing transitions between them.   In this picture, spindown rates and X-ray states, along with radio nulling, drifts, and mode changes,  are all {\em collateral} effects.  
Discrete states are tied to the physics of the magnetic polar cap (PC) while switching 
involves both local and global effects.  The suitability of simple Markov processes for
describing the distributions of state durations in most cases suggests that state switching 
is largely  stochastic.   This is most easily understood as  self-driven stochasticity
of the acceleration region.  The natural time scale
for such variations $\lesssim 1~\mu s$.  However,  the stochastic resonance model
for the quasi-periodicity in B1931+24 suggests the action of
 a forcing function with a 
long period $\sim 38$~days that is easiest 
to understand in terms of modulation of the
return current from  an equatorial disk outside the light cylinder. 
Equatorial disks
are a common feature in recent discussions of magnetospheres
 and in numerical simulations of pulsar magnetospheres 
 \citep[][]{1985A&A...144...72K, 2005A&A...442..579C, 2010MNRAS.408L..41T,
 2012ApJ...749....2K, 2012ApJ...746...60L, 2012ApJ...746L..24L}
 and have long
been argued for by \citet[][]{1980Ap&SS..72..175M}.  
Dramatic changes of  magnetospheric structure by pulsar-disk interactions were a conclusion
of \citet[][]{2009ARep...53...86B}.
Recent magnetosphere models incorporate a  `Y' point where
current from a disk flows along the separatrix to both magnetic PCs
\citep[e.g.][]{2005A&A...442..579C, 2006MNRAS.368.1055T}.    This feature may provide 
the most direct feedback from a pulsar's environment to  its PCs.


\subsection{Circuit Analogs}

The NS surface and atmosphere at the PC and the acceleration region above  have  been likened to the idealized unidirectional current flow of a diode
\citep[e.g.][]{1979ApJ...231..854A, 2008ApJ...688..499T}.   The 
analogy has been extended to  include
resistivity, capacitance, and inductance in {\em linear} circuits
\citep[e.g.][]{1991ApJ...378..239S, 2001ApJ...547..959J, 2006ChJAA...6..217X}
that, in principle, could show oscillations with well-defined time scales. 
However, nonlinear circuits can have metastable states with broad, Markov-type
durations when state switching occurs from noise-like voltage variations. 
A nonlinear, real-world  diode with a forward  bias voltage, a back-biased
capacitance, and a finite recovery time to  a reversal from forward to backward  
bias can   show universal chaotic behavior when driven sinusoidally \citep[][]{rh82}.
Nonlinear circuits are amenable to a Markov description
 when switching between states is rapid compared
to the residence time in any state and when `microscopic' dynamics within a state can be
ignored compared to the step-size of the Markov chain \citep[e.g.][]{2012JChPh.136q4119B}. 
 An asymmetric  Schmitt trigger, which features hysteresis and nonlinearity, has unequal  switching thresholds for
positive and negative inputs.   It shows Markov variations for a random-noise input  but can  show stochastic resonance when  driven sinusoidally, behavior that is generic to any asymmetric bistable system  \citep[][]{m+99}.  


NS and their magnetospheres differ substantially from laboratory circuits  because voltage drops and current flows along field lines  are not controlled but instead  are dynamical quantities, providing a  richer variety of possible phenomena.  

Pulsar magnetospheres strive to be force free ($\Evec\cdot\Bvec = 0$)
by filling themselves with 
space charge at the Goldreich-Julian (GJ) density
$\rhoGJ \approx -{\Omegavec \cdot \Bvec}/{2\pi c}$, where $\Omegavec$ is the spin vector.
Standard models for pulsars include a corotating magnetosphere with 
$\rho = \rhoGJ$ and an  open field line (OFL) region
where charge is lost through a current density $J \approx c \rhoGJ $,
but $\rho = \rhoGJ$   almost everywhere except in  acceleration regions where 
$\Evec\cdot\Bvec \ne 0$.  The potential drop sustains the current flow that
contributes to the spindown torque and, along with secondary particles,  produces radio to gamma-ray emission.    
The OFL region
and magnetic PC are  defined self-consistently from
  the combined ambient stellar magnetic field  and  current-driven field; 
  the latter,  in turn, depends on boundary conditions
imposed by the global structure of the magnetosphere \citep[e.g.][]{2012ApJ...749....2K}. 
 
Recent simulations \citep{2012ApJ...749....2K, 2012ApJ...746...60L, 2012ApJ...746L..24L} 
demonstrate exquisitely  how the  spindown rate  depends on 
the magnetic flux threading the light cylinder and takes on limiting values 
for  vacuum ($\rho = 0$) and force-free ($\rho = \rhoGJ$) conditions in the OFL region.
 Spindown rates in these limits differ by a factor of four for inclination angles of 45$^{\circ}$, easily bracketing  the
2.5:1 range seen in long-term intermittent pulsars.  For active, steady pulsars the magnetosphere appears to find a configuration that lies between the force-free
and vacuum solutions. However, there is no obvious mechanism
for defining dual (let alone multiple) states in what appears to be a continuum for the resistivity,
which is a free parameter in determining the type of magnetosphere 
\citep[][]{2012ApJ...746...60L, 2012ApJ...746L..24L}. 

We attribute discrete current states  to  changes in the
relative contributions from ions, protons, electrons, and $e^{\pm}$ pairs. 
This situation applies to  
an $\Omegavec \cdot \Bvec <0$ geometry for  the OFL region at 
 the NS surface,  the situation considered by  \citet[][]{1975ApJ...196...51R} and  \citet[][]{2011MNRAS.414..759J, 2012MNRAS.423.3502J, 2013arXiv1302.5260J},
 that requires positive charges above the PC. 
The opposite case with $\Omegavec \cdot \Bvec > 0$   has comparatively 
uncomplicated electron flows that may not allow pair production
and  radio emission.  However, if it does,  state changes are not expected. 
The $\Omegavec \cdot \Bvec <0$
and $\Omegavec \cdot \Bvec > 0$ cases may then account for 
the stark differences between pulsars having similar $P$ and $\dot P$. 
Positive charges  include ions with a wide range
of charge to mass ratios resulting from particle showers onto the polar cap
\citep[][]{2012MNRAS.423.3502J} along with protons and any positrons produced in
pair avalanches.     One form of current self-regulation involves thermionic emission
from a PC that is kept hot by backflowing electrons from $e^{\pm}$ pairs
\citep[][]{1980ApJ...235..576C}. Small changes in temperature can change the ion
current density $J_i$ exponentially if the current flow is not  space-charge limited (SCL). 
In this case, cessation of $e^{\pm}$ production allows the PC to cool and $J_i$ will 
decrease.   Unless another charge source can compensate, the voltage drop will increase
while the spindown rate decreases.   Coherent radiation that requires
a substantial pair plasma  will also shut off.    With SCL flow, however,
cessation of pair production will cause a radio null but without a large decrease in 
torque on the star.

\subsection{Return Currents and Stellar Charge}

Current outflow must be matched by a return current, at least on average, 
if the NS is not to become charged sufficiently to halt current outflow from the polar cap.   
While a disaster for explaining {\em steady} pulsar radiation, episodic charging and discharging 
is a plausible mechanism for intermittency because it could take little time (about one spin
period) to switch between charge states.    
 In some models the return current is from separated 
$e^{\pm}$ pairs created within the magnetosphere, while in others it originates from 
an equatorial disk exterior to the magnetosphere.       If a mismatch develops
between outflow and inflow, it will self regulate if pair production is robust
as in high-field, fastly spinning pulsars.  However, longer-period or
weak-field pulsars that operate near the threshold for pair production,
will  have a higher propensity for state changes.  It is therefore notable
that fast states (classical nulling,  drifting and mode changes), long-term intermittency,
discrete torques, and X-ray states have been seen only in pulsars with $P\gtrsim 0.3$~s
and are common for $P\gtrsim 1$~s. 
Furthermore, nulling is most visible in objects with small angles between the spin axis
and magnetic moment \citep[][]{1992ApJ...394..574B,  2008ApJ...682.1152C}.   Given
both of these conditions, it is plausible that currents originating from an equatorial disk
can drive the system into and out of a state of dormancy, at least as far as radio emission
is concerned.  

Return currents from an exterior disk, if highly episodic,  
will modulate activity at the PC as well as
 alter the structure of the magnetosphere \citep[][]{2009ARep...53...86B}. 
 Depending on their constituency, they can also 
 `dope' the NS surface with  ions that differ from those that drift to  the top of the thin 
 atmosphere from below \citep[][]{2011MNRAS.414..759J}, which  
 are fractionated
 in charge-to-mass ratio by  backflow  electrons  \citep[][]{2012MNRAS.423.3502J}.
 In laboratory diodes with non-SCL current flow, the current density
 depends on the type of material as well as on the voltage drop  \citep[][]{gbs04}.    
 A similar effect may
 occur in pulsars where the charge carriers could be a state-dependent mixture
 of surface and backflow ions.  
Like any circuit, the current flow in the magnetosphere can halt altogether.    
With sufficient net charge, the combined electrostatic and induced 
 potential drops can shut off the conventional PC accelerator by reverse biasing the system
 \citep[][]{2002A&A...384..414P}. 

These considerations suggest that if return currents originate from or are affected by
a disk external to the light cylinder, the electrodynamics of the disk itself will be involved 
in state-change phenomena.    Disk instabilities  combined with perturbations from
contaminating material
\citep[e.g. asteroids;][]{2008ApJ...682.1152C}
 may introduce a wide range of time scales, including orbital
time scales.   Some of these may be stochastic and others
periodic.   In addition, magnetic reconnection within the disk from
magnetic fields with oppositive polarities above and below the disk can inject relativistic
particles into the magnetosphere \cite[e.g.][]{2005A&A...442..579C, 2006MNRAS.367...19K, 2006MNRAS.368.1055T}.

\subsection{Discrete Drift States}

We associate fast state
 changes  with
 the acceleration region just above the NS surface and atmosphere
 e.g. \citep[][]{1975ApJ...196...51R}.  
 Fast nulling and pulse-shape modes  suggest alteration of  the
particle composition in current density  and the number of secondary
$e^{\pm}$ pairs.  
Preferred drift rates may be associated with changing numbers of
sub-beams in a rotating carousel.  The number of sub-beams is likely involved
in determining the local  charge difference $\Delta\rho = \rho - \rhoGJ$
and thus would affect the  $\Evec\times\Bvec$ drift velocity. 
The number of carousel beams $\sim 20$ in 
B0943+10 \citep[][]{1999ApJ...524.1008D} 
while another object, B1822-09, appears to
  involve only two subpulse beams in B mode
and three in Q mode \citep[][]{2012MNRAS.427..180L}.  Sub-beam
creation and annihilation would be an attractive process for producing discrete
values of subpulse drifts and other state attributes. However,    two pulsars mentioned earlier
(B0031$-$07 and B2319+60) have approximately constant  $P_2$ 
in the three  observed drift modes in which there is an  
 increase in $P_3$ and a decrease in the drift rate $P_2/P_3$.   
 The spatial separation of sub-beams is directly related to $P_2$ suggesting,
 at least for these two objects,  a constant number
of sub-beams because $N_{sb} \propto P_4/P_2$.   In other pulsars the variability
of $P_2$ is not well constrained.  
An alternative to the rotating carousel model
  is the diocotron instability that has been suggested as a cause of subpulse drifts \cite[][]{2006A&A...445..779F}, which  results from
differential angular rotation across the OFL region.  Magnetospheric models that incorporate
this instability are not well-enough  developed to provide a basis for further discussion in
this paper. 

\subsection{Long-lived States, Energetics, and MSPs}

The  range of  time scales seen in  pulsar intermittency is large so it is natural to ask
 what other time scales are relevant and what the energetics are.   
Given that observed time scales 
have little relationship to characteristic times of NS and their magnetospheres and that
processes exterior to the magnetosphere may be involved, objects with much longer
state durations (e.g. years, decades)  plausibly exist that will be discovered in
comprehensive radio surveys that have a large product of time coverage and field-of-view. 
\citep[e.g.][]{2010PASA...27..272M, 2013PASA...30....6M}.
   
As for energetics,  it is well known that pulsar radio luminosities are
only small fractions ($\sim 10^{-8}$ to $10^{-2}$) of the spin-down loss rates $\dot E$
but  can be comparable to particle energy fluxes.  It therefore follows that  sporadic on-state emission can be seen over distances that are limited by the amount of energy available
to a burst of particle acceleration. If  magnetospheres make occasional transitions from 
a near vacuum state to a force-free state and back, the available energy can be much greater than for typical radio pulsars and the resulting pulses may be visible over extragalactic 
distances.   Such pulses might be related to any that may be seen with dispersion measures too large to be accounted for by sources embedded in the Galactic disk
\citep[e.g.][]{2011MNRAS.415.3065K}.

\section{Summary and Conclusions}
\label{sec:summary}

In this paper it has been shown that metastable states in pulsar phenomena are common
and that while many cases involve binary states, some pulsars display multiple states.  
Pulsars are well known to show pulse profiles that are peculiar to each object but are
consistent over tens of years or more.   Similarly, the presence and properties of state
switching also appear to have stationary statistics.     These properties include the 
mean durations  and the long-term frequencies of occurrence of
particular states, including the nulling fraction often used to characterize objects that
show intensity nulls.    Collectively, the phenemona suggest that states are defined by long-lasting features of neutron stars and their magnetospheres.  In the Markov interpretation
presented in this paper, the transition matrix appears to be epoch independent in most
objects other than B1931+24. 

We have also shown that  quasi-periodic switching on time scales of weeks for B1931+24 is consistent with a Markov system that
displays stochastic resonance.   Stochastic resonance is seen in systems that
are driven by both stochastic fluctuations and a forcing function.   Most studies assume
the forcing function is sinusoidal, but arbitrary deterministic or stochastic
forcings are also generally possible.  For B1931+24,  the observed quasi-periodicity allows
either a deterministic or quasi-periodic forcing function.  It is not obvious how to 
discriminate between the two observationally.   One possibility is to monitor
this object (and other long-term intermittent pulsars) with  high-cadence observations     so that switching statistics can be better compared with  specific models. 

To date, only a small fraction of pulsars have been studied with sufficient S/N to allow 
quantitative evaluation of a Markov (or other) model.      Even for many of the
well-studied objects, pulse-to-pulse variations cause overlap of on-and-off-state
intensities, leading to false positives from algorithms that identify state changes.
Future work with now-available wideband receiver systems on existing telescopes
and eventually with new array telescopes (ASKAP, MeerKAT, and the SKA) can 
improve the discrimination between states as well as expand the sample of objects
that can be studied in this way.  

Of equal interest are objects that do not show (or have not yet shown) any kind of state
change.  Pulsars with periods $\lesssim 0.3$~s, including MSPs,   are among these.
Such pulsars may have such robust pair production that they are not susceptible to
state changes because return currents are self-generated inside the magnetosphere rather
than originating from an external disk. 
In addition, there are objects that do not show state changes yet are in the same region
of the $P$-$\dot P$ diagram as pulsars  that do.  
It is important to establish more firmly the occurrence
or non-occurence of metastable states in these pulsars.   If pulsars that are otherwise similar 
in $P$ and $\dot P$ show markedly different variability,
additional factors need to be considered that may be peculiar to individual objects.   
Candidate factors include the angular offset of the magnetic axis from the spin axis, higher-order multiples of the magnetic field, surface composition, 
and environmental effects in or near the equatorial disk.  Another possibility
is that the charge polarity at the PC, given by the sign of $\Omegavec \cdot \Bvec$,
determines whether a pulsar displays metastable states.

Future work  that can illuminate some of the issues raised in this work include: 
\begin{enumerate}
\item
Simultaneous observations over wide frequency ranges to test whether some null states represent redistribution of radio flux to a different frequency band;
\item
Long-term monitoring campaigns with single-pulse sampling to identify state changes
that occur on both short and long time scales;
\item Characterization of state changes as Markov processes in more pulsars and testing
for long-term stationarity of the transition matrix; 
\item Detailed investigations of single pulses from MSPs to assess whether state changes
occur that are qualitatively similar to those of canonical pulsars with $\sim 10^{12}$~G fields;
\item
 Higher sensitivity observations that  discriminate between null and burst states
 with minimal or no false transitions; these can also clarify whether there are true
 null states  or simply low-radio-emission states; 
\item
High cadence timing measurements that allow differentiation between state changes and
other contributions to spin noise;
\item Very low frequency observations to test whether there is coherent ion emission;
and
\item Additional simultaneous high-energy and radio observations to elucidate the 
nature of emission states and the locations of the emission regions for the different
wavebands.

 \end{enumerate}

Acknowledgements:  
I thank Shami Chatterjee,  Ira Wasserman, Dusty Madison, Robert Wharton, and Joanna Rankin for helpful discussions during the writing of this paper; 
Ramesh Bhat for providing data used in Figure 1;
Jon Arons for numerous discussions in the past about the puzzle of long-term pulsar variability; Curt Michel for his sustained efforts in  educating the pulsar community on the existence of magnetospheric configurations that are `dead' observationally;
and Michael Kramer and Andrew Lyne for early discussions about B1931+24. 

\newpage
{

}


\begin{thebibliography}{27}
\setlength{\itemsep}{-0.5mm}
\expandafter\ifx\csname natexlab\endcsname\relax\def\natexlab#1{#1}\fi

\bibitem[Anishchenko, Anufrieva \& Vadivasova(2006)]{a+06}
Anishchenko, V. S., Anufrieva, M. V., \& Vadivasova, T. E. 2006,
Technical Physics Letters, 32, 873 

\bibitem[Arathi, Rajasekar, \& Kurths(2011)]{a+11}
Arathi, S., Rajasekar, S., \& Kurths, J. 2011, International Journal of Bifurcation and Chaos,
21, 2729





\bibitem[Arons 
\& Scharlemann(1979)]{1979ApJ...231..854A} Arons, J., \& Scharlemann, E.~T.\ 1979, \apj, 231, 854 


\bibitem[Backer(1973)]{1973ApJ...182..245B} Backer, D.~C.\ 1973, \apj, 182, 
245

\bibitem[Barsukov et al.(2009)]{2009ARep...53...86B} Barsukov, D.~P., 
Polyakova, P.~I., \& Tsygan, A.~I.\ 2009, Astronomy Reports, 53, 86

\bibitem[Bartel et al.(1982)]{1982ApJ...258..776B} Bartel, N., Morris, D., 
Sieber, W., \& Hankins, T.~H.\ 1982, \apj, 258, 776 

\bibitem[Becker 
\& Rein ten Wolde(2012)]{2012JChPh.136q4119B} Becker, N.~B., \& Rein ten Wolde, P.\ 2012, \jcp, 136, 174119

\bibitem[Bhat et 
al.(2007)]{2007A&A...462..257B} Bhat, N.~D.~R., Gupta, Y., Kramer, M., et al.\ 2007, \aap, 462, 257 






\bibitem[Biggs(1992)]{1992ApJ...394..574B} Biggs, J.~D.\ 1992, \apj, 394, 
574 



\bibitem[Burke-Spolaor et al.(2012)]{2012MNRAS.423.1351B} Burke-Spolaor, 
S., Johnston, S., Bailes, M., et al.\ 2012, \mnras, 423, 1351 

\bibitem[Camilo et al.(2012)]{2012ApJ...746...63C} Camilo, F., Ransom, 
S.~M., Chatterjee, S., Johnston, S., \& Demorest, P.\ 2012, \apj, 746, 63 


\bibitem[Chen et al.(2011)]{2011ApJ...741...48C} Chen, J.~L., Wang, H.~G., 
Wang, N., et al.\ 2011, \apj, 741, 48 

\bibitem[Cheng 
\& Ruderman(1980)]{1980ApJ...235..576C} Cheng, A.~F., \& Ruderman, M.~A.\ 1980, \apj, 235, 576

\bibitem[Cognard et al.(1996)]{1996ApJ...457L..81C} Cognard, I., Shrauner, 
J.~A., Taylor, J.~H., \& Thorsett, S.~E.\ 1996, \apjl, 457, L81



\bibitem[Contopoulos(2005)]{2005A&A...442..579C} Contopoulos, I.\ 2005, \aap, 442, 579 

\bibitem[Cordes(1979)]{1979SSRv...24..567C} Cordes, J.~M.\ 1979, \ssr, 24, 
567



\bibitem[Cordes(1983)]{1983AIPC..101...98C} Cordes, J.~M.\ 1983, 
Positron-Electron Pairs in Astrophysics, 101, 98


\bibitem[Cordes 
\& Downs(1985)]{1985ApJS...59..343C} Cordes, J.~M., \& Downs, G.~S.\ 1985, \apjs, 59, 343

\bibitem[Cordes et al.(2004)]{2004ApJ...612..375C} Cordes, J.~M., Bhat, 
N.~D.~R., Hankins, T.~H., McLaughlin, M.~A., 
\& Kern, J.\ 2004, \apj, 612, 375 

\bibitem[Cordes 
\& Shannon(2008)]{2008ApJ...682.1152C} Cordes, J.~M., \& Shannon, R.~M.\ 2008, \apj, 682, 1152

\bibitem[Deich et al.(1986)]{1986ApJ...300..540D} Deich, W.~T.~S., Cordes, 
J.~M., Hankins, T.~H., \& Rankin, J.~M.\ 1986, \apj, 300, 540 

\bibitem[Deshpande 
\& Rankin(1999)]{1999ApJ...524.1008D} Deshpande, A.~A., \& Rankin, J.~M.\ 1999, \apj, 524, 1008 



\bibitem[Deneva et al.(2009)]{2009ApJ...703.2259D} Deneva, J.~S., Cordes, 
J.~M., McLaughlin, M.~A., et al.\ 2009, \apj, 703, 2259 


\bibitem[Esamdin et al.(2012)]{2012ApJ...759L...3E} Esamdin, A., Abdurixit, 
D., Manchester, R.~N., \& Niu, H.~B.\ 2012, \apjl, 759, L3 





\bibitem[Fung et al.(2006)]{2006A&A...445..779F} Fung, P.~K., Khechinashvili, D., \& Kuijpers, J.\ 2006, \aap, 445, 779


\bibitem[Gajjar et al.(2012)]{2012MNRAS.424.1197G} Gajjar, V., Joshi, 
B.~C., \& Kramer, M.\ 2012, \mnras, 424, 1197 



\bibitem[Gammaitoni et al.(1998)]{g+98}
Gammaitoni, L., H\"anggi, P., Junk, P., Marchesoni, F. 1998, Rev. Mod. Phys.,
70, 224



\bibitem[Gil et al.(2006)]{2006ApJ...650.1048G} Gil, J., Melikidze, G., 
\& Zhang, B.\ 2006, \apj, 650, 1048 


\bibitem[Goncalves, Barroso, \& Sandonato(2004)]{gbs04}
Goncalves, J. A. N., Barroso, J. J., \& Sandonato, G. M. 2004,
Diamond and Related Materials, 13, 60. 



\bibitem[Groth(1975)]{1975ApJS...29..443G} Groth, E.~J.\ 1975, \apjs, 29, 
443 

\bibitem[Harding et al.(2002)]{2002ApJ...576..366H} Harding, A.~K., 
Muslimov, A.~G., \& Zhang, B.\ 2002, \apj, 576, 366

\bibitem[Harding 
\& Muslimov(2011)]{2011ApJ...743..181H} Harding, A.~K., \& Muslimov, A.~G.\ 2011, \apj, 743, 181 

\bibitem[Hermsen et al.(2013)]{h+13}
Hermsen, W. et al. 2013, Science, 339, 436

\bibitem[Hibschman 
\& Arons(2001)]{2001ApJ...554..624H} Hibschman, J.~A., \& Arons, J.\ 2001, \apj, 554, 624 




\bibitem[Huguenin et al.(1970)]{1970ApJ...162..727H} Huguenin, G.~R., 
Taylor, J.~H., \& Troland, T.~H.\ 1970, \apj, 162, 727 


\bibitem[Janssen 
\& van Leeuwen(2004)]{2004A&A...425..255J} Janssen, G.~H., \& van Leeuwen, J.\ 2004, \aap, 425, 255 

\bibitem[Jessner et al.(2001)]{2001ApJ...547..959J} Jessner, A., Lesch, H., 
\& Kunzl, T.\ 2001, \apj, 547, 959 


 

\bibitem[Jones(1982)]{1982MNRAS.200.1081J} Jones, P.~B.\ 1982, \mnras, 200, 
1081 

\bibitem[Jones(1986)]{1986MNRAS.222..577J} Jones, P.~B.\ 1986, \mnras, 222, 
577 

\bibitem[Jones(2011)]{2011MNRAS.414..759J} Jones, P.~B.\ 2011, \mnras, 414, 
759 

\bibitem[Jones(2012)]{2012MNRAS.423.3502J} Jones, P.~B.\ 2012, \mnras, 423, 
3502

\bibitem[Jones(2013)]{2013arXiv1302.5260J} Jones, P.\ 2013, arXiv:1302.5260

\bibitem[Kalapotharakos et al.(2012)]{2012ApJ...749....2K} Kalapotharakos, 
C., Kazanas, D., Harding, A., \& Contopoulos, I.\ 2012, \apj, 749, 2 

\bibitem[Karastergiou et al.(2011)]{2011MNRAS.415..251K} Karastergiou, A., 
Roberts, S.~J., Johnston, S., et al.\ 2011, \mnras, 415, 251 


\bibitem[Keane et al.(2011)]{2011MNRAS.415.3065K} Keane, E.~F., Kramer, M., 
Lyne, A.~G., Stappers, B.~W., \& McLaughlin, M.~A.\ 2011, \mnras, 415, 3065 

\bibitem[Komissarov(2006)]{2006MNRAS.367...19K} Komissarov, S.~S.\ 2006, 
\mnras, 367, 19




\bibitem[Kloumann 
\& Rankin(2010)]{2010MNRAS.408...40K} Kloumann, I.~M., \& Rankin, J.~M.\ 2010, \mnras, 408, 40 

\bibitem[Kramer et al.(2006)]{2006Sci...312..549K} Kramer, M., Lyne, A.~G., 
O'Brien, J.~T., Jordan, C.~A., \& Lorimer, D.~R.\ 2006, Science, 312, 549

\bibitem[Krause-Polstorff 
\& Michel(1985)]{1985A&A...144...72K} Krause-Polstorff, J., \& Michel, F.~C.\ 1985, \aap, 144, 72 



\bibitem[Latham et al.(2012)]{2012MNRAS.427..180L} Latham, C., Mitra, D., 
\& Rankin, J.\ 2012, \mnras, 427, 180 



\bibitem[Li et al.(2012b)]{2012ApJ...746...60L} Li, J., Spitkovsky, A., 
\& Tchekhovskoy, A.\ 2012, \apj, 746, 60 

\bibitem[Li et al.(2012c)]{2012ApJ...746L..24L} Li, J., Spitkovsky, A., 
\& Tchekhovskoy, A.\ 2012, \apjl, 746, L24 


\bibitem[Lorimer et al.(2012)]{2012ApJ...758..141L} Lorimer, D.~R., Lyne, 
A.~G., McLaughlin, M.~A., et al.\ 2012, \apj, 758, 141

\bibitem[Lundgren et al.(1995)]{1995ApJ...453..433L} Lundgren, S.~C., 
Cordes, J.~M., Ulmer, M., et al.\ 1995, \apj, 453, 433


\bibitem[Lyne 
\& Ashworth(1983)]{1983MNRAS.204..519L} Lyne, A.~G., \& Ashworth, M.\ 1983, \mnras, 204, 519 


\bibitem[Lyne et al.(2010)]{2010Sci...329..408L} Lyne, A., Hobbs, G., 
Kramer, M., Stairs, I., \& Stappers, B.\ 2010, Science, 329, 408 

\bibitem[Marchesoni, Apostolico \& Santucci(1999)]{m+99}
Marchesoni, F., Apostolico, F. \& Santucci, S. 1999, Phys. Rev. E, 59, 3958


\bibitem[McLaughlin et al.(2006)]{2006Natur.439..817M} McLaughlin, M.~A., 
Lyne, A.~G., Lorimer, D.~R., et al.\ 2006, \nat, 439, 817

\bibitem[Macquart et al.(2010)]{2010PASA...27..272M} Macquart, J.-P., 
Bailes, M., Bhat, N.~D.~R., et al.\ 2010, \pasa, 27, 272

\bibitem[Medin \& Lai(2007)]{2007MNRAS.382.1833M} Medin, Z., \& Lai, D.\ 2007, \mnras, 382, 1833 


\bibitem[Michel(1980)]{1980Ap&SS..72..175M} Michel, F.~C.\ 1980, \apss, 72, 175 


\bibitem[Michel(1991)]{m91}
Michel, F. C., Theory of neutron star magnetospheres, University of Chicago Press, Chicago, IL, USA, 1991a.

\bibitem[Michel(2010)]{2010HEAD...11.1621M} Michel, F.~C.\ 2010, Bulletin 
of the American Astronomical Society, 42, 682 

\bibitem[Murphy et al.(2013)]{2013PASA...30....6M} Murphy, T., Chatterjee, 
S., Kaplan, D.~L., et al.\ 2013, \pasa, 30, 6 


\bibitem[Palliyaguru et al.(2011)]{2011MNRAS.417.1871P} Palliyaguru, N.~T., 
McLaughlin, M.~A., Keane, E.~F., et al.\ 2011, \mnras, 417, 1871 

\bibitem[Papoulis(1991)]{Papoulis91}
Papoulis, A. 1991, ``Probability, Random Variables, and Stochastic Processes,''
(McGraw-Hill, New York), pp. 638-639

\bibitem[P{\'e}tri et 
al.(2002)]{2002A&A...384..414P} P{\'e}tri, J., Heyvaerts, J., \& Bonazzola, S.\ 2002, \aap, 384, 414 

\bibitem[Rabiner(1989)]{r89}
Rabiner, L. R. 1989, Proc. IEEE, 77, 257


\bibitem[Rankin 
\& Stappers(2008)]{2008AIPC..983..112R} Rankin, J.~M., \& Stappers, B.\ 2008, 40 Years of Pulsars: Millisecond Pulsars, Magnetars and More, 983, 112 

\bibitem[Redman et al.(2005)]{2005MNRAS.357..859R} Redman, S.~L., Wright, 
G.~A.~E., \& Rankin, J.~M.\ 2005, \mnras, 357, 859 

\bibitem[Redman 
\& Rankin(2009)]{2009MNRAS.395.1529R} Redman, S.~L., \& Rankin, J.~M.\ 2009, \mnras, 395, 1529 

\bibitem[Ritchings(1976)]{1976MNRAS.176..249R} Ritchings, R.~T.\ 1976, 
\mnras, 176, 249 

\bibitem[Rollins \& Hunt(1982)]{rh82}
Rollins, R. W. \& Hunt, E. R. 1982, PRL, 49, 1295

\bibitem[Ruderman 
\& Sutherland(1975)]{1975ApJ...196...51R} Ruderman, M.~A., \& Sutherland, P.~G.\ 1975, \apj, 196, 51 
 

\bibitem[{{Shannon} \& {Cordes}(2010)}]{2010ApJ...725.1607S}
{Shannon}, R.~M., \& {Cordes}, J.~M. 2010, \apj, 725, 1607

\bibitem[Shibata(1991)]{1991ApJ...378..239S} Shibata, S.\ 1991, \apj, 378, 
239 

\bibitem[Smith et al.(2001)]{2001MNRAS.322..209S} Smith, I.~A., Michel, 
F.~C., \& Thacker, P.~D.\ 2001, \mnras, 322, 209 










\bibitem[Thompson(2008)]{2008ApJ...688..499T} Thompson, C.\ 2008, \apj, 
688, 499 


\bibitem[Timokhin(2006)]{2006MNRAS.368.1055T} Timokhin, A.~N.\ 2006, 
\mnras, 368, 1055

\bibitem[Timokhin(2010)]{2010MNRAS.408L..41T} Timokhin, A.~N.\ 2010, 
\mnras, 408, L41 






\bibitem[van Leeuwen 
\& Timokhin(2012)]{2012ApJ...752..155V} van Leeuwen, J., \& Timokhin, A.~N.\ 2012, \apj, 752, 155 

\bibitem[van Leeuwen et 
al.(2002)]{2002A&A...387..169V} van Leeuwen, A.~G.~J., Kouwenhoven, M.~L.~A., Ramachandran, R., Rankin, J.~M., \& Stappers, B.~W.\ 2002, \aap, 387, 169



\bibitem[Vivekanand 
\& Joshi(1997)]{1997ApJ...477..431V} Vivekanand, M., \& Joshi, B.~C.\ 1997, \apj, 477, 431 

\bibitem[Wang 
\& Hirotani(2011)]{2011ApJ...736..127W} Wang, R.-B., \& Hirotani, K.\ 2011, \apj, 736, 127

\bibitem[Wang et al.(2007)]{2007MNRAS.377.1383W} Wang, N., Manchester, 
R.~N., \& Johnston, S.\ 2007, \mnras, 377, 1383 


\bibitem[Weltevrede et 
al.(2006)]{2006A&A...445..243W} Weltevrede, P., Edwards, R.~T., \& Stappers, B.~W.\ 2006, \aap, 445, 243 






\bibitem[Wright 
\& Fowler(1981a)]{1981IAUS...95..211W} Wright, G.~A.~E., \& Fowler, L.~A.\ 1981a, Pulsars: 13 Years of Research on Neutron Stars, 95, 211

\bibitem[Wright 
\& Fowler(1981b)]{1981A&A...101..356W} Wright, G.~A.~E., \& Fowler, L.~A.\ 1981b, \aap, 101, 356 

\bibitem[Xu et al.(2006)]{2006ChJAA...6..217X} Xu, R.-X., Cui, X.-H., 
\& Qiao, G.-J.\ 2006, \cjaa, 6, 217

\bibitem[Young et al.(2013)]{2013MNRAS.tmp..449Y} Young, N.~J., Stappers, 
B.~W., Lyne, A.~G., et al.\ 2013, \mnras, 449 


\bibitem[Young et al.(2012)]{2012MNRAS.427..114Y} Young, N.~J., Stappers, 
B.~W., Weltevrede, P., Lyne, A.~G., \& Kramer, M.\ 2012, \mnras, 427, 114






\end{thebibliography}
\end{document}